\documentclass[fleqn,usenatbib]{aa}
\usepackage{txfonts}
\usepackage{graphicx}

\usepackage{natbib}
\usepackage{longtable}
\usepackage{subeqnarray}
\usepackage{cases}
\usepackage[colorlinks=true,citecolor=blue]{hyperref}
\usepackage{amsmath}
\usepackage[switch]{lineno} 
\usepackage{multirow}
\usepackage{caption}

\newcommand{\orcid}[1]{\href{https://orcid.org/#1}{\includegraphics[width=10pt]{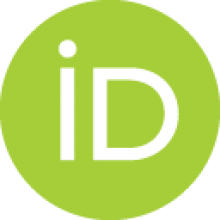}}}

\begin{document}

\title{Kinetic temperature of massive star-forming molecular clumps measured with formaldehyde}
\subtitle{VI. The photodissociation region M17SW}

\author{{X. Zhao\orcid{0009-0005-6436-0966}\inst{\ref{inst1},\ref{inst2}}}
\and X. D. Tang\orcid{0000-0002-4154-4309}\inst{\ref{inst1},\ref{inst2},\ref{inst3}}
\and C. Henkel\orcid{0000-0002-7495-4005}\inst{\ref{inst4},\ref{inst1}}
\and K. M. Menten\inst{\ref{inst4}}\thanks{In memoriam.}
\and Y. Gong\orcid{0000-0002-3866-414X}\inst{\ref{inst5}}
\and Y. Sun\orcid{0000-0002-3904-1622}\inst{\ref{inst5}}
\and Y. P. Ao\orcid{0000-0003-3139-2724}\inst{\ref{inst5}}
\and T. Liu\orcid{0000-0002-5286-2564}\inst{\ref{inst6}}
\and X. Lu\orcid{0000-0003-2619-9305}\inst{\ref{inst6}}
\and D. L. Li\inst{\ref{inst1},\ref{inst2},\ref{inst3}}
\and Y. X. He\orcid{0000-0002-8760-8988}\inst{\ref{inst1},\ref{inst2},\ref{inst3}}
\and K. Wang\orcid{0000-0002-7237-3856}\inst{\ref{inst7}}
\and X. P. Chen\orcid{0000-0003-3151-8964}\inst{\ref{inst5}}
\and J. W. Wu\orcid{0000-0001-7808-3756}\inst{\ref{inst8},\ref{inst2}}
\and J. Esimbek\orcid{0000-0003-4910-1390}\inst{\ref{inst1},\ref{inst2},\ref{inst3}}
\and J. J. Zhou\orcid{0000-0003-0356-818X}\inst{\ref{inst1},\ref{inst2},\ref{inst3}}
\and X. W. Zheng\inst{\ref{inst9}}
\and J. J. Qiu\orcid{0000-0002-9829-8655}\inst{\ref{inst10}}
\and J. S. Li\orcid{0009-0008-8664-5681}\inst{\ref{inst1},\ref{inst2}}
\and C. S. Luo\orcid{0009-0005-5281-5716}\inst{\ref{inst1},\ref{inst2}}
\and Q. Zhao\orcid{0009-0006-2956-0783}\inst{\ref{inst1},\ref{inst2}}
\and L. D. Liu\orcid{0009-0001-1882-1291}\inst{\ref{inst1},\ref{inst2}}
\and C. Y. Wang\inst{\ref{inst1},\ref{inst2}}
}

\titlerunning{Kinetic temperatures in M17SW}
\authorrunning{Zhao et al.}

\institute{
State Key Laboratory of Radio Astronomy and Technology, Xinjiang Astronomical Observatory, CAS, Urumqi 830011, PR China\label{inst1}\\
\email{tangxindi@xao.ac.cn}
\and University of Chinese Academy of Sciences, Beijing 100080, PR China \label{inst2}
\and Xinjiang Key Laboratory of Radio Astrophysics, Urumqi 830011, PR China \label{inst3}
\and Max-Planck-Institut f\"{u}r Radioastronomie, Auf dem H\"{u}gel 69, 53121 Bonn, Germany \label{inst4}
\and Purple Mountain Observatory, Chinese Academy of Sciences, Nanjing 210008, PR China \label{inst5}
\and Shanghai Astronomical Observatory, Chinese Academy of Sciences, 80 Nandan Road, Shanghai 200030, PR China \label{inst6}
\and Kavli Institute for Astronomy and Astrophysics, Peking University, Beijing 100871, PR China \label{inst7}
\and National Astronomical Observatories, Chinese Academy of Sciences, Beijing 100101, PR China \label{inst8}
\and School of Astronomy and Space Science, Nanjing University, Nanjing 210093, PR China \label{inst9}
\and School of Mathmatics and Physics, Jinggangshan University, Ji’an 343009, PR China \label{inst10}
}

\abstract
{The kinetic temperature structure of the photodissociation region M17SW was mapped using the IRAM 30\,m telescope. This mapping employed the para-H$_2$CO triplet ($J_{\rm K_aK_c}$\,=\,3$_{03}$--2$_{02}$, 3$_{22}$--2$_{21}$, 
and 3$_{21}$--2$_{20}$) near 218\,GHz on a scale of $\sim$0.2\,pc. The kinetic temperatures were derived by modeling the average H$_2$CO line ratios (3$_{22}$--2$_{21}$/3$_{03}$--2$_{02}$\,+\,3$_{21}$--2$_{20}$/3$_{03}$--2$_{02}$) with the RADEX non-local thermodynamic equilibrium approach. These temperatures range from 28 to 181\,K with an average of 54.2\,$\pm$\,0.3\,K at a spatial density of 5.5$\times$10$^{5}$\,cm$^{-3}$.
Comparing with the temperature measurements obtained from multiple transitions of NH$_3$\,(1,1)--(6,6) and the far infrared (FIR) dust continuum, the H$_2$CO lines show temperatures similar to those measured by NH$_3$ but slightly higher than values derived from FIR observations. The high kinetic temperatures observed from H$_2$CO are associated with the ultracompact H\,{\scriptsize II} region UC1, dense clumps, as well as H$_2$O and CH$_3$OH masers, showing a similar distribution as NH$_3$. This indicates that dense gas in the M17SW region is heated by star formation activity.
The presence of a significant gas temperature gradient across the M17SW region, as measured by H$_2$CO and NH$_3$, provides direct evidence for gas heated predominantly by radiation emitted from the OB star cluster NGC\,6618.
On a smaller scale, the dense gas surrounding the dense clumps experiences significant heating from internal protostars and/or young stellar objects. Higher temperatures traced by H$_2$CO are linked to turbulence on a scale of $\sim$0.2\,pc.
The complex temperature structure of the M17SW region is revealed by H$_2$CO and NH$_3$, which may be attributed to both large-scale external radiative heating and small-scale internal radiative and turbulent heating.
}

\maketitle

\section{Introduction}
\label{Sect:Introduction}
The physical properties of the interstellar medium (ISM) significantly 
influence the star formation rate, spatial distribution, and fundamental characteristics of subsequent 
generations of stars, including their initial mass function and the elemental composition
(e.g., \citealt{Paumard2006,Kennicutt1998a,Kennicutt1998b,Klessen2007,Papadopoulos2011,Zhang2018,Zhang2025,Tang2019,Beuther2025,Gong2025}). 
Massive stars play a crucial role in shaping their environments and influencing subsequent star formation 
through feedback such as outflows, stellar winds, ultraviolet (UV) radiation, and gas compression by expanding H\,{\scriptsize II} regions. However, the details of 
the massive star formation process and the ways in which their feedback modifies the initial conditions 
for massive stars remain inadequately understood. Radiative and mechanical interactions between massive stars
and molecular clouds are fundamental for understanding galaxy evolution. Massive stars influence the 
evolution of molecular clouds by eroding and photo-evaporating their surfaces through intense UV radiation 
fields (e.g., \citealt{Hernandez-Vera2023,Gong2026}). Additionally, UV radiation plays a crucial role in establishing 
the thermal gas pressure within star-forming clouds, with effects that can extend across various spatial scales, 
from the edges of molecular clouds to entire star-forming galaxies. Due to their powerful radiation fields 
and stellar winds, massive stars can either trigger or suppress the formation of new stars by compressing 
or disrupting their natal clouds (e.g., \citealt{Elmegreen1977,Walch2012,Kim2018b,Pabst2019,Luisi2021,Hernandez-Vera2023}). 
Therefore, it is essential to study the physical conditions of molecular clouds exposed to stellar UV radiation fields, 
particularly in the context of photo-dissociation or photo dominated regions (PDRs), to evaluate the effects of stellar feedback.
The thermal balance of the gas in such regions is governed by a combination of radiative and turbulent heating processes. 
In PDRs and cloud surfaces, the photoelectric effect on dust grains is the dominant radiative heating mechanism (e.g., \citealt{Bakes1994,Weingartner2001}),
followed by the photodissociation of H$_2$ and its UV pumping (e.g., \citealt{Hollenbach1991,Girichidis2020}).
In denser regions, dust grains heated by the ambient radiation field transfer energy to the gas particles through collisions, efficiently coupling the dust and gas temperatures. 
Turbulent heating, on the other hand, arises from the dissipation of supersonic turbulence through shocks (e.g., \citealt{Guesten1988,Pan2009,Ao2013}).

M17, also known as the Omega Nebula, is one of the most extensively studied young, active massive star-forming 
regions and ranks among the brightest H\,{\scriptsize II} regions in the sky \citep{Lada1974,Hanson1997}. 
Within this H\,{\scriptsize II} region lies the young open cluster NGC\,6618, which comprises approximately 
100 OB stars \citep{Beetz1976,Chini1980,Hanson1997,Chini2000,Hoffmeister2008}. These stars contribute to the
heating of the surrounding gas and ionization of the M17 region \citep{Lada1991,Hoffmeister2006}.
NGC\,6618 has physically segmented the infrared structure of M17 into two distinct areas: M17 North and 
M17 South \citep{Lada1991,Lada1976}. In the southwestern portion of M17, a compact molecular cloud known as 
M17SW has been identified. Extensive observations indicate that M17SW is one of the most representative PDRs 
in the Milky Way \citep{Wilson1999,Brogan2001,Pellegrini2007}, with a distance of $\sim$2\,kpc \citep{Xu2011}. 
The molecular cloud M17SW is situated adjacent to the H\,{\scriptsize II} region within M17, prompting extensive
studies of M17SW as a PDR across high-energy, optical, infrared, and radio wavelengths (e.g., \citealt{Felli1984,Mundy1987,Stutzki1988,Hobson1992,Hobson1993,Jiang2002,Chen2012,Chen2013,Tang2013,Tang2014,Lim2020,Liu2022,Yanza2022,Zhu2023,Zhao2026}). 
The heating and compression effects from the adjacent H\,{\scriptsize II} region may induce the collapse 
of the M17SW region, potentially leading to the formation of a new generation of stars. This makes M17SW a
prime observational site for studying star formation processes 
(e.g., \citealt{Chini1980,Lemke1981,Felli1984,Stutzki1990,Hanson1997,Nielbock2001,Reid2006,Perez-Beaupuits2012,Chen2021,Chen2025,Yin2022,Zhou2024}).

The thermal structure of the M17SW region has been a subject of extensive study (see Appendix\,\ref{Sect:Previous} for 
a synthesis of previous results). Previous observations indicate that the M17SW region is deeply influenced by the UV radiation from the H\,{\scriptsize II} region, resulting in a decreasing temperature trend away from the radiating source, and the direction is from northeast to southwest (e.g., \citealt{Gatley1979,Thronson1983,Harris1987,Stutzki1988,Guesten1988,Stutzki1990,Greaves1992,Meixner1992,Bergin1994,Perez-Beaupuits2010}). 
Specifically, the dense molecular gas and dust at the interface between the H\,{\scriptsize II} region and the molecular cloud are heated primarily by UV radiation from the external OB cluster NGC\,6618 (e.g., \citealt{Gatley1979,Stutzki1990,Meixner1992}). This external heating scenario has been further supported by subsequent molecular line and infrared continuum studies (e.g., \citealt{Perez-Beaupuits2010, Nishimura2018, Hoang2022}). Evidence of dense gas being heated by turbulence in the M17SW region was discovered by \cite{Guesten1988} through measurements using NH$_3$ on a scale of approximately 0.4\,pc.

These studies provide valuable insights into the various layers of molecular clouds and the gas and dust 
temperature variations within the PDR of M17SW at different spatial scales.
However, these studies are fundamentally limited by their angular resolutions. 
This lack of resolving power introduces significant uncertainty regarding the small-scale physics;
the observed gradient could represent a smooth trend or be composed of discrete, unresolved temperature fluctuations. 
Consequently, a comprehensive observations detailing the temperature of dense gas related to star formation in the M17SW region, 
particularly at a sub-pc scale and with high resolution, remains lacking. Therefore, further research and 
high-resolution observations of M17SW are essential for a complete characterization of the thermal 
properties of dense gas.

Formaldehyde (H$_{2}$CO) is a ubiquitous molecule in the interstellar medium (ISM), and its abundance 
remains stable across various evolutionary stages of star-forming regions. Previous observations 
indicate that the abundance of H$_{2}$CO does not vary by more than an order of magnitude throughout 
different stages of star formation 
(e.g., \citealt{Mangum1990,Mangum1993,Caselli1993,Johnstone2003,Gerner2014,Tang2017a,Tang2017b,Tang2018b,Liu2020,Gong2023}). 
H$_{2}$CO exhibits numerous spectral transitions, ranging from centimeter to submillimeter wavelengths,
and extensive observational studies demonstrate that its spectral lines can be employed to trace 
the physical properties of molecular clouds
(e.g., \citealt{Henkel1980,Henkel1983,Mangum1993,Mangum2008,Mangum2013a,Mangum2019,Ginsburg2011,Ginsburg2016,Ao2013,Tang2017a,Tang2017b,Tang2018a,Tang2018b,Tang2021,Immer2016,Zhao2024}).
The relative populations of the $K_{\rm a}$ ladders of H$_{2}$CO are primarily governed by collisions, 
making the ratios of H$_{2}$CO line fluxes involving different $K_{\rm a}$ ladders effective indicators 
of temperature \citep{Mangum1993,Tang2018b,Mazumdar2022,Kahle2023}. 
The ratios of those H$_{2}$CO transitions have been utilized to measure the temperature of dense gas in various contexts, 
including Galactic star formation regions 
(e.g., \citealt{Mangum1990,Mangum1993,Caselli1993,Johnstone2003,Gerner2014,Tang2017a,Tang2017b,Tang2018b,Liu2020,Khan2026}), 
the Galactic center  \citep{Qin2008,Ao2013,Johnston2014,Ginsburg2016,Immer2016,Lu2017,Zhang2025,Yue2026}, 
and external galaxies \citep{Muhle2007,Tang2017b,Tang2021,Mangum2019}.
Previous observations indicate that H$_{2}$CO is a sensitive tracer of dense gas in well-known Galactic PDRs, 
such as the Orion Bar, the Horsehead Nebula, and M8 (e.g., \citealt{Leurini2010,Guzman2011,Cuadrado2017,Tang2018a,Kahle2024}).
Additionally, H$_{2}$CO is useful in studying low-metallicity environments with strong UV radiation found 
in the Magellanic Clouds (e.g., \citealt{Heikkila1999,Wang2009,Tang2017b,Tang2021,Henkel2022}). 
It is arguably one of the most effective molecular probes available for analyzing dense gas in PDRs. 
The distributions of multiple H$_2$CO spectral lines have been mapped in the M17SW region at low resolution 
\citep{Lada1975,Gardner1981,Mundy1987,Tang2013,Tang2014}.

\begin{figure*}[t]
\centering
\includegraphics[width=1.0\textwidth,angle=0]{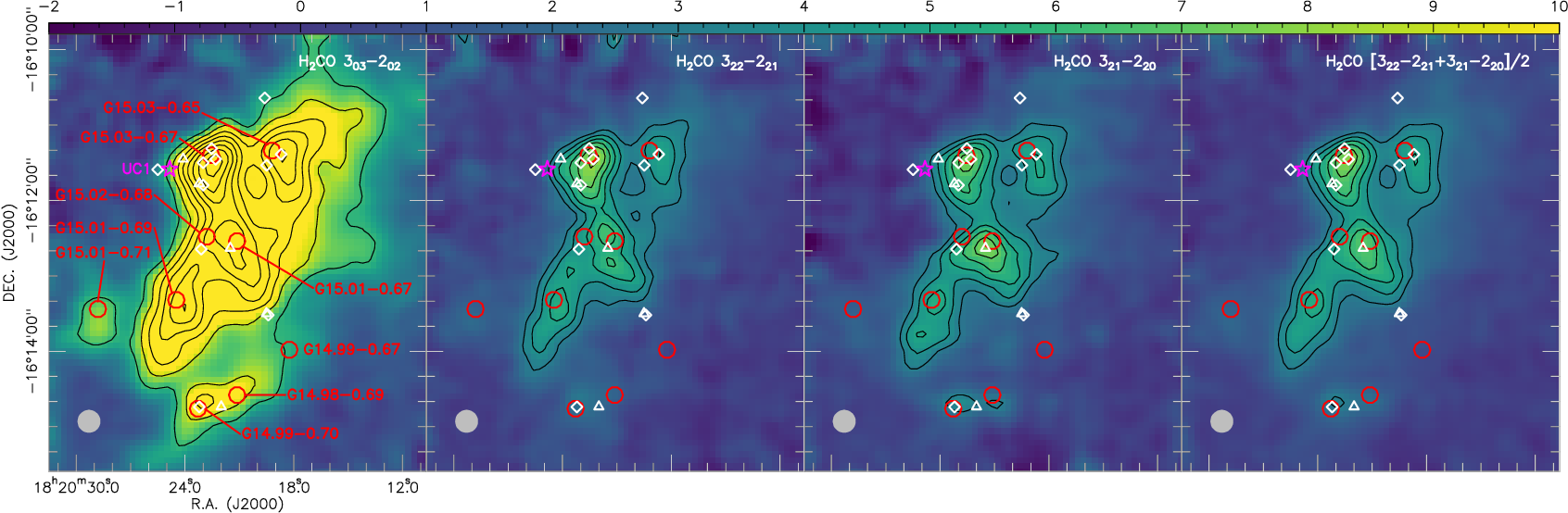}
\caption{Velocity-integrated intensity maps of H$_{2}$CO\,3$_{03}$--2$_{02}$ (\emph{left}),
3$_{22}$--2$_{21}$ (\emph{mid-left}), 3$_{21}$--2$_{20}$ (\emph{mid-right}), 
and combined 3$_{22}$--2$_{21}$ and 3$_{21}$--2$_{20}$ (\emph{right}) are shown on a $T_{\rm mb}$ scale,
with the color bar in units of K\,km\,s$^{-1}$. These maps are integrated over the velocity range of 
$V_{\rm LSR}$\,=\,14 to 26\,km\,s$^{-1}$ toward M17SW. Contour levels are from 6.0 to 27.0\,K\,km\,s$^{-1}$ 
with steps of 3.0\,K\,km\,s$^{-1}$ for H$_{2}$CO\,3$_{03}$--2$_{02}$, from 3 to 10\,K\,km\,s$^{-1}$ with steps 
of 1.0\,K\,km\,s$^{-1}$ for 3$_{22}$--2$_{21}$, 3$_{21}$--2$_{20}$, and combined 3$_{22}$--2$_{21}$ and 3$_{21}$--2$_{20}$.
The red circles represent the location of dense clumps identified from 850\,$\mu$m continuum emission by \cite{Eden2019}.
The white diamonds and triangles and the magenta star indicate the locations of 22\,GHz H$_{2}$O masers 
\citep{Forster1999,Breen2010,Breen2011,Urquhart2011}, Class\,I CH$_3$OH masers \citep{Kurtz2004,Gomez2010,Kim2018a,Yang2020,Yang2023},
and the ultra-compact H\,{\scriptsize II} region UC1 \citep{Felli1984}, respectively.
The beam size of each image is shown in the lower left corner.}
\label{fig:H2CO-intensity-maps}
\end{figure*}

The main goal of this paper is to map the kinetic temperature ($T_{\rm kin}$) structure of the PDR M17SW
at a scale of $\sim$0.2\,pc by utilizing the 
para-H$_{2}$CO triplet ($J_{\rm K_aK_c}$\,=\,3$_{03}$--2$_{02}$, 3$_{22}$--2$_{21}$, and 3$_{21}$--2$_{20}$), 
which has a critical density on the order of 
$n_{\rm crit}$(H$_2$CO\,3$_{03}$--2$_{02}$)\,$\sim$\,a few\,10$^{5}$\,cm$^{-3}$ \citep{Shirley2015}. 
We also aim to investigate the gas heating mechanisms affecting the dense gas in the context of star formation.
The structure of the paper is as follows: In Sects.\,\ref{Sect:Observations-and-data-reduction} and \ref{Sect:Results}, 
we present our observations of the H$_{2}$CO triplet, data reduction, and results. The resulting kinetic 
temperatures derived from H$_2$CO are discussed in Sect.\,\ref{Sect:Discussion}. Finally, we summarize our 
main conclusions in Sect.\,\ref{Sect:Sumamry}. This paper is part of the "Kinetic temperature of massive
star-forming molecular clumps measured with formaldehyde" series of studies exploiting H$_2$CO as a probe 
of gas conditions in a variety of Galactic and extragalactic sources.

\section{Observations and data reduction}
\label{Sect:Observations-and-data-reduction}
We used the 9 dual-polarization pixel HEterodyne Receiver Array (HERA, \citealt{Schuster2004}) of the IRAM\,30\,m telescope\footnote{\tiny Based on observations obtained with the IRAM\,30\,m
telescope. IRAM is supported by INSU/CNRS (France), MPG (Germany), and IGN (Spain).} during May 2017 and October 2018
to map the M17SW molecular cloud. The H$_{2}$CO triplet ($J_{\rm K_aK_c}$\,=\,3$_{03}$--2$_{02}$, 3$_{22}$--2$_{21}$, 
and 3$_{21}$--2$_{20}$) transitions have rest frequencies of 218.222, 218.475, and 218.760\,GHz, respectively.
The central frequency was set to 218.475 GHz to ensure simultaneous measurements of all three lines.
Within $\sim$1\,GHz bandwidth, the backend spectrometer has 5377 channels, with each channel having a width of
$\sim$0.27\,km\,s$^{-1}$. The main beam size at $\sim$218\,GHz is about 12$''$. The linear scale is $\sim$0.12\,pc at 
a distance of  2\,kpc. The main beam efficiency and the forward hemisphere
efficiency\footnote{\tiny https://publicwiki.iram.es/Iram30mEfficiencies} 
are 0.60 and 0.94, respectively. The observations were performed in the on-the-fly (OTF) model and were centered on 
$\alpha_{2000}$\,=\,18$^{\rm h}$20$^{\rm m}$21\hbox{$\,.\!\!^{\rm s}$}1 
and $\delta_{2000}$\,=\,-16$\degr$11$'$35\hbox{$\,.\!\!^{\prime\prime}$}7.
Four $\sim$3\,$\times$\,3\,arcmin$^2$ maps were measured with steps of 3.5$''$ in both right ascension and declination,
and the mapped area is $\sim$4.7\,$\times$\,5.1\,arcmin$^2$ ($\sim$2.7\,$\times$\,2.9\,pc$^{2}$).

Data was reduced with GILDAS\footnote{\tiny http://www.iram.fr/IRAMFR/GILDAS}.
To enhance the signal-to-noise ratio (S/N) in individual channels, we smoothed three adjacent spectral channel, 
to a velocity resolution $\sim$0.8\,km\,s$^{-1}$.
The OTF maps were smoothed to a spatial resolution of 18$''$ and a 6$''$ grid using a Gaussian convolution kernel with inverse variance weighting.
The comprehensive dataset comprising 5413 individual data points, each one corresponding to a distinct spectrum including various transitions.
The detection rate was $\sim$41\% for H$_{2}$CO\,3$_{03}$--2$_{02}$, and $\sim$23\% for the 
H$_{2}$CO\,3$_{22}$--2$_{21}$ and 3$_{21}$--2$_{20}$ lines, each exhibiting S/N of $\gtrsim$3$\sigma$.
All H$_{2}$CO spectra with S/Ns above 3$\sigma$ were fitted by Gaussians.
A typical rms noise level (1$\sigma$) of H$_{2}$CO\,3$_{03}$--2$_{02}$ is $\sim$0.12\,K ($T_{\rm mb}$ scale) 
at a velocity resolution of $\sim$0.8\,km\,s$^{-1}$.

The H$_{2}$CO\,3$_{22}$--2$_{21}$ and 3$_{21}$--2$_{20}$ transitions exhibit analogous distributions and line profiles 
(brightness temperature, line width, and velocity in our observations, see Fig.\,\ref{fig:H2CO-intensity-maps}), 
analogous to the situation in other star forming regions
(e.g., \citealt{Mangum1993,Tang2017a,Tang2017b,Tang2018a,Tang2018b,Tang2021,Zhao2024,Zhang2025}).
They belong to the same $K_{\rm a}$ ladder, and the energies between the ground state and these excited states 
are almost identical, $E_{\rm u}$\,$\simeq$\,68\,K. Typically, H$_{2}$CO\,3$_{03}$--2$_{02}$ shows stronger emissions 
than H$_{2}$CO\,3$_{22}$--2$_{21}$ and 3$_{21}$--2$_{20}$. Furthmore, the line ratios of 
H$_{2}$CO\,3$_{22}$--2$_{21}$/3$_{03}$--2$_{02}$ and 3$_{21}$--2$_{20}$/3$_{03}$--2$_{02}$ demonstrate analogous 
behaviors to probe the kinetic temperature of dense gas \citep{Mangum1993,Tang2017a}.
Hence, we combined the line intensities of the two transitions channel by channel to further improve the S/Ns of 
H$_{2}$CO\,3$_{22}$--2$_{21}$ and 3$_{21}$--2$_{20}$.

The H$_{2}$CO\,3$_{03}$--2$_{02}$ spectral lines reveal multiple distinct velocity components at certain 
locations within the M17SW molecular cloud. 
These multiple velocity components originate in the northern edge of M17SW and sequentially connect through G15.03-0.65, 
G15.01-0.67, G15.02-0.68, and the region around G14.98–0.69 (see Fig.\,\ref{fig:H2CO-intensity-maps}).
We combined the data from these two velocity components based on their integrated intensities ratio (details see Sect.\,\ref{Sect:T&NT-thermal}). 
Therefore, the gas temperature in positions with dual velocity components was not measured separately. 
Other positions were modeled using a single Gaussian. The resultant line width measurements may 
be overestimated in these regions.

\section{Results}
\label{Sect:Results}
\subsection{Overview}
\label{Sect:Overview}
The velocity-integrated intensity distributions of the H$_2$CO triplet and the combined H$_{2}$CO\,3$_{22}$--2$_{21}$
and 3$_{21}$--2$_{20}$ lines in the PDR M17SW ($V_{\rm LSR}$\,=\,14 to 26\,km\,s$^{-1}$) are 
depicted in Fig.\,\ref{fig:H2CO-intensity-maps}. Figures\,\ref{fig:H2CO-moment1&2} and \ref{fig:H2CO-channel} 
display the intensity-weighted velocity field (moment 1), line width (moment 2), and channel maps of 
H$_{2}$CO\,3$_{03}$--2$_{02}$. The observed H$_2$CO spectra of nine dense clumps, identified by 850\,$\mu$m 
continuum emission \citep{Eden2019} are illustrated in Fig.\,\ref{fig:observed_clumps}. 
The positions were converted to offset coordinates relative to 
our phase center at $\alpha_{2000}$\,=\,18$^{\rm h}$20$^{\rm m}$21\hbox{$\,.\!\!^{\rm s}$}1 
and $\delta_{2000}$\,=\,-16$\degr$11$'$35\hbox{$\,.\!\!^{\prime\prime}$}7. 
These positions were then registered to the nearest pixel in our data cube to ensure alignment with our observational grid.
The locations of these dense clumps in M17SW are marked in Fig.\,\ref{fig:H2CO-intensity-maps} and detailed in 
Table\,\ref{table:Clump-Parameters}. Table\,\ref{table:Clump-Parameters} also presents the parameters of 
the Gaussian fits for these dense clumps, encompassing velocity-integrated intensity ($\int T_{\rm mb}$\,d$v$), 
local standard of rest velocity ($V_{\rm LSR}$), full width at half maximum (FWHM) line width, 
and peak temperature ($T_{\rm mb}$) of the H$_2$CO spectra.

\subsection{Distribution of H$_{2}$CO}
\label{Sect:distribution-H2CO}
The H$_{2}$CO\,3$_{03}$--2$_{02}$ velocity integrated intensity map shows the extensive distribution and provides a 
clear clumpy dense structure of M17SW  (see the top left panel of Fig.\,\ref{fig:H2CO-intensity-maps}).
The strong H$_{2}$CO emission is consistent with the dense dust clumps \citep{Eden2019}. The spatial distribution of H$_{2}$CO\,3$_{03}$--2$_{02}$ also closely aligns with previous observational
results utilizing other gas tracers, such as C$^{18}$O\,(2--1) and CO\,(6--5 and 7--6)
\citep{Stutzki1990,Perez-Beaupuits2010}, as well as dust emission at wavelengths of 450--850\,$\mu$m
(e.g.,\,\citealt{Hobson1993,Chen2025}). In contrast, H$_{2}$CO\,3$_{22}$--2$_{21}$ and 3$_{21}$--2$_{20}$ exhibit more
localized distributions, primarily detectable in the densest regions of M17SW, as depicted in the top right and bottom 
two panels of Fig.\,\ref{fig:H2CO-intensity-maps}.

\begin{figure}[t]
\centering
\includegraphics[width=0.46\textwidth]{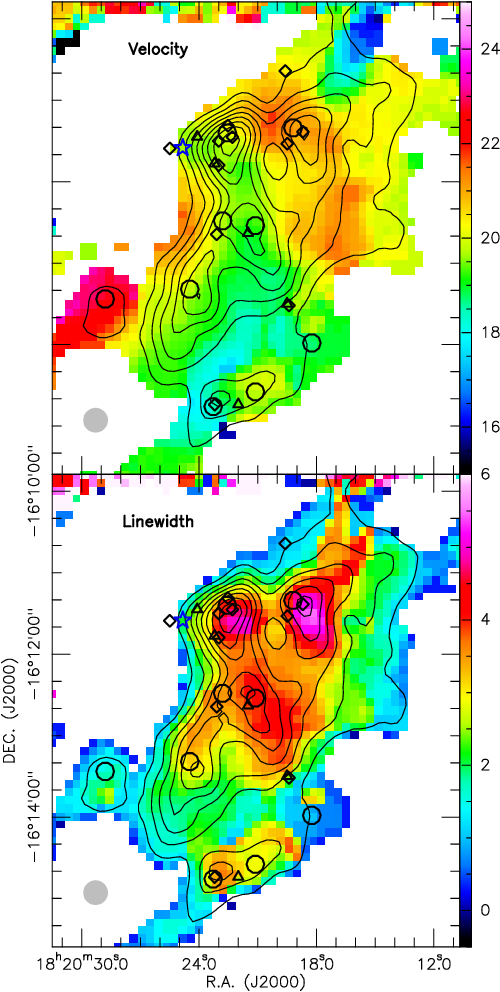}
\caption{The intensity-weighted velocity field (moment 1, \emph{top}) and line width (moment 2, \emph{bottom}) 
of H$_{2}$CO\,3$_{03}$--2$_{02}$ are presented. The unit of each colour bar is km\,s$^{-1}$. 
Black contours show the integrated intensity of H$_{2}$CO\,3$_{03}$--2$_{02}$ (same as in Fig.\,\ref{fig:H2CO-intensity-maps}).
The symbols are the same as Fig.\,\ref{fig:H2CO-intensity-maps}.}
\label{fig:H2CO-moment1&2}
\end{figure}

The intensity-weighted velocity (moment\,1) and channel maps of H$_{2}$CO\,(3$_{03}$--2$_{02}$) elucidate the intricate
velocity field within M17SW, as depicted in Figs.\,\ref{fig:H2CO-moment1&2} and \ref{fig:H2CO-channel}. Velocity components
ranging from approximately 20 to 23\,km\,s$^{-1}$ are discernible at specific locations, including the clump G15.01-0.71,
the eastern periphery of clump G15.02-0.68, clump G15.03-0.65 and its southern extended region, the connecting bridge between
clumps G15.03-0.67 and G15.03-0.65, and the northwestern sector of M17SW. Additionally, velocity components within the
range of approximately 16 to 18\,km\,s$^{-1}$ are observed in the northern and southern sectors of M17SW. Notably, the 
eastern dense region of the primary structure of M17SW exhibits a velocity of around 19\,km\,s$^{-1}$.

\begin{figure*}[t]
\centering
\includegraphics[width=0.95\textwidth]{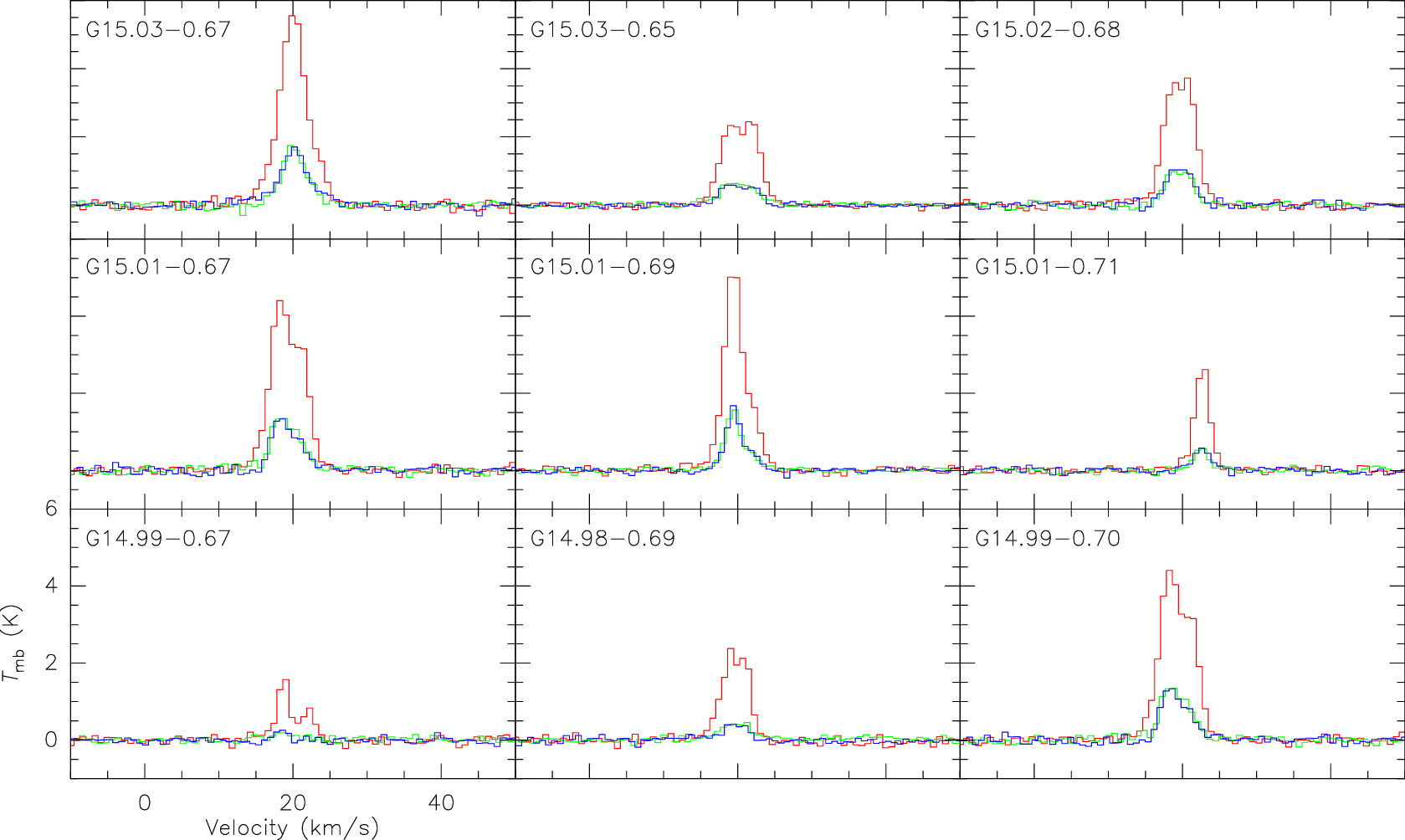}
\caption{Observed H$_{2}$CO spectra of the dense clumps in M17SW. Red, green, and blue lines show 
H$_{2}$CO\,3$_{03}$--2$_{02}$, 3$_{21}$--2$_{20}$, and 3$_{22}$--2$_{21}$, respectively. 
The dense clumps were identified from 850\,$\mu$m continuum emission by \cite{Eden2019} 
(details see Sect.\,\ref{Sect:Overview} and Table\,\ref{table:Clump-Parameters}). }
\label{fig:observed_clumps}
\end{figure*}

\begin{table*}[h]
\scriptsize
\caption{Parameters of the dense clumps in M17SW.}
\centering
\begin{tabular}
{ccclccccccccccccccc}
\hline\hline
Clump & Offset &Transition &$\int T_{\rm mb}$\,d$v$ &$V_{\rm LSR}$ &FWHM &$T_{\rm mb}$ & $\sigma_{\rm T}$ & $\sigma_{\rm NT}$ & $c_{\rm s}$  & $\mathcal{M}$ & $T_{\rm kin}$ & $T_{\rm dust}^{b}$ \\
       & " , " &  &K\,km\,s$^{-1}$ &km\,s$^{-1}$ &km\,s$^{-1}$ &K &km\,s$^{-1}$ & km\,s$^{-1}$ & km\,s$^{-1}$ & & K & K\\
\hline
G15.03-0.67        &   18, 12    &3$_{03}$--2$_{02}$                         &26.4 (0.3) &20.0 (0.1) &4.6 (0.1) &5.4 & 0.14 (0.01) & 1.96 (0.03) & 0.50 (0.01) & 4.0 & 71 (3) & 35.9   \\ 
                     &             &(3$_{21}$--2$_{20}$+3$_{22}$--2$_{21}$)/2  &7.5 (0.1)  &20.0 (0.1) &4.4 (0.1) &1.6 &             &             &             &     &              \\ 
G15.03-0.65        & --30, 18    &3$_{03}$--2$_{02}$                         &20.2 (0.2) &21.1 (0.1) &6.3 (0.1) &3.0 & 0.12 (0.01) & 2.67 (0.03) & 0.44 (0.01) & 6.1 & 55 (2) & 32.7   \\ 
                     &             &(3$_{21}$--2$_{20}$+3$_{22}$--2$_{21}$)/2  &4.9 (0.1)  &20.9 (0.1) &5.9 (0.2) &0.8 &             &             &             &     &              \\ 
G15.02-0.68$^{a}$  &   24, --54  &3$_{03}$--2$_{02}$                         &18.8 (2.1) &19.7 (0.1) &2.9 (0.2) &6.2 & 0.13 (0.02) & 1.21 (0.07) & 0.45 (0.08) & 2.7 & 58 (20) & --   \\ 
                     &             &(3$_{21}$--2$_{20}$+3$_{22}$--2$_{21}$)/2  &4.7 (1.0)  &19.4 (0.2) &2.6 (0.3) &1.7 &             &             &             &     &              \\ 
G15.01-0.67        &   0, --54   &3$_{03}$--2$_{02}$                         &23.9 (0.2) &19.2 (0.1) &5.3 (0.1) &4.2 & 0.13 (0.01) & 2.25 (0.03) & 0.47 (0.01) & 4.8 & 63 (3) & 31.5  \\ 
                     &             &(3$_{21}$--2$_{20}$+3$_{22}$--2$_{21}$)/2  &6.3 (0.2)  &18.8 (0.1) &4.5 (0.1) &1.3 &             &             &             &     &              \\ 
G15.01-0.69        & 48, --102   &3$_{03}$--2$_{02}$                         &18.4 (0.2) &19.6 (0.1) &3.6 (0.1) &4.9 & 0.13 (0.01) & 1.51 (0.02) & 0.47 (0.01) & 3.2 & 62 (2) & --    \\ 
                     &             &(3$_{21}$--2$_{20}$+3$_{22}$--2$_{21}$)/2  &4.8 (0.1)  &19.4 (0.1) &3.1 (0.1) &1.4 &             &             &             &     &              \\ 
G15.01-0.71        & 108, --108  &3$_{03}$--2$_{02}$                         &6.9 (0.1)  &22.7 (0.1) &2.4 (0.1) &2.7 & 0.12 (0.01) & 1.03 (0.02) & 0.42 (0.02) & 2.4 & 51 (4) & 38.7  \\ 
                     &             &(3$_{21}$--2$_{20}$+3$_{22}$--2$_{21}$)/2  &1.6 (0.1)  &22.5 (0.1) &2.7 (0.2) &0.6 &             &             &             &     &              \\ 
G14.99-0.67        & --42, --144 &3$_{03}$--2$_{02}$                         &5.6 (0.2)  &19.4 (0.1) &5.0 (0.3) &1.1 & 0.09 (0.01) & 2.10 (0.12) & 0.31 (0.01) & 6.7 & 28 (2) & --   \\ 
                     &             &(3$_{21}$--2$_{20}$+3$_{22}$--2$_{21}$)/2  &0.6 (0.1)  &18.4 (0.2) &2.2 (0.4) &0.2 &             &             &             &     &              \\ 
G14.98-0.69$^{a}$  & 0, --180    &3$_{03}$--2$_{02}$                         &10.3 (0.7) &19.6 (0.1) &3.4 (0.2) &3.2 & 0.13 (0.02) & 1.46 (0.09) & 0.44 (0.06) & 3.3 & 56 (17) & --   \\ 
                     &             &(3$_{21}$--2$_{20}$+3$_{22}$--2$_{21}$)/2  &2.5 (0.4)  &18.7 (0.5) &6.7 (1.1) &0.5 &             &             &             &     &              \\ 
G14.99-0.70        &   30, --192 &3$_{03}$--2$_{02}$                         &12.0 (0.2) &18.2 (0.1) &5.9 (0.1) &1.9 & 0.12 (0.01) & 2.52 (0.05) & 0.43 (0.01) & 5.9 & 53 (3) & --   \\ 
                     &             &(3$_{21}$--2$_{20}$+3$_{22}$--2$_{21}$)/2  &2.8 (0.1)  &18.0 (0.1) &6.4 (0.3) &0.4 &             &             &             &     &              \\ 
 
\hline
\end{tabular}
\label{table:Clump-Parameters}
\tablefoot{Offsets relative to our reference position for M17SW (see Sect.\,\ref{Sect:Overview} 
and Fig.\,\ref{fig:H2CO-intensity-maps}, \citealt{Eden2019}). The table also lists the velocity integrated intensity, $\int T_{\rm mb}$\,d$v$, local standard of rest velocity, $V_{\rm LSR}$, full width
at half maximum line width (FWHM), and peak temperature ($T_{\rm mb}$). These parameters were obtained from Gaussian fitting using CLASS as part of the GILDAS 
software package. 
For the thermal and nonthermal line widths, $\sigma_{\rm T}$ and $\sigma_{\rm NT}$, the sound velocity $c_{\rm s}$, 
the Mach number $\mathcal{M}$, the kinetic temperature $T_{\rm kin}$, and the dust temperature $T_{\rm dust}$,
see Sects.\,\ref{Sect:Results} and \ref{Sect:Turbulent-heating}.
Values in parenthesis are 1$\sigma$ uncertainties.
$^{(a)}$Clumps G15.02-0.68 and G14.98-0.69 are located in a region with two velocity components, 
their parameters are the combined values of both components (see Sect.\,\ref{Sect:Observations-and-data-reduction}).
$^{(b)}$The dust temperature taken from \citealt{Urquhart2018}.}
\end{table*}

The line width distribution (moment\,2) is illustrated in the bottom panel of Fig.\,\ref{fig:H2CO-moment1&2}. 
It is evident that regions exhibiting high line widths, exceeding approximately 3\,km\,s$^{-1}$, are closely 
linked to dense clumps, specifically G15.03-0.67, G15.03-0.65, G15.02-0.68, G15.01-0.67, and G15.01-0.69. 
Moderate line widths (2--3\,km\,s$^{-1}$) predominantly encircle the dense clumps within M17SW, while lower 
line widths ($\lesssim$2\,km\,s$^{-1}$) are observed at the periphery of the primary structure of M17SW and 
at clump G15.01-0.71. The fitted line width values for H$_{2}$CO\,3$_{03}$--2$_{02}$ range from 
1.2 to 7.5\,km\,s$^{-1}$, with an average value of 3.8$\pm$0.1\,km\,s$^{-1}$ (errors reported represent standard 
deviations of the mean).
Approximately 34\% of spectra with high line widths ($\geq$3\,km\,s$^{-1}$) exhibit dual velocity components in the fitted data, 
while this proportion is 53\% for those with moderate line widths (2--3\,km\,s$^{-1}$).

\begin{figure*}[t]
\centering
\includegraphics[width=1.0\textwidth]{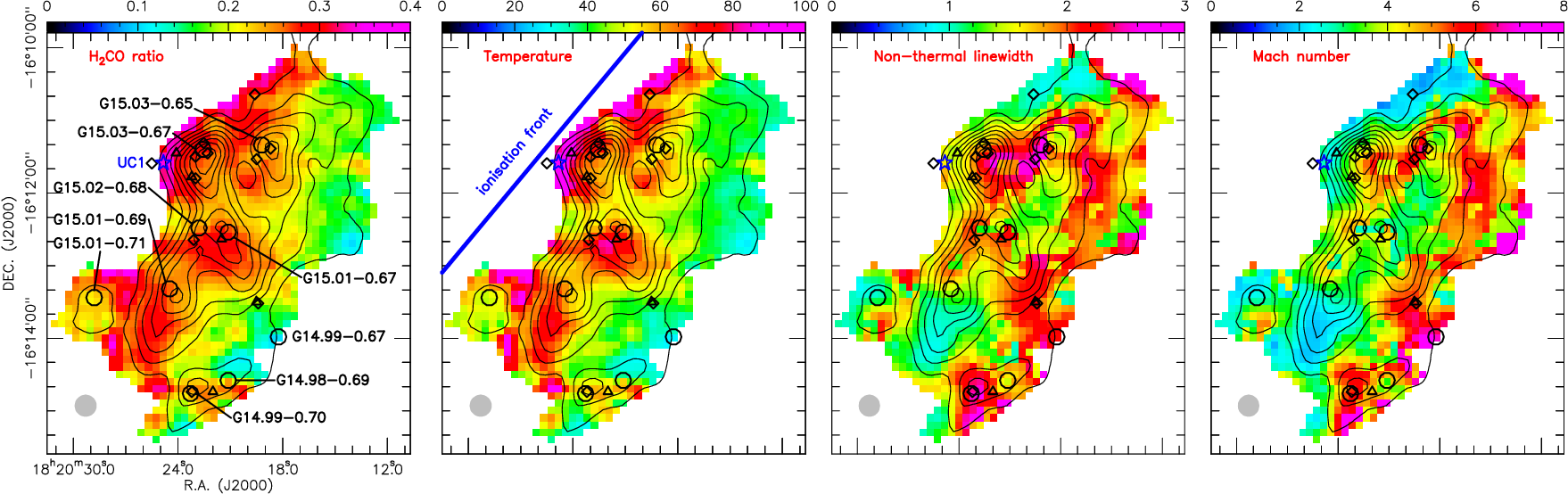}
\caption{Parameter maps of M17SW.
\emph{Left}: the averaged velocity-integrated intensity ratio map of H$_{2}$CO, calculated as
0.5$\times$[(3$_{22}$--2$_{21}$\,+\,3$_{21}$--2$_{20})$/3$_{03}$--2$_{02}$] (see Sect.\,\ref{Sect:H2CO-Ratio}). 
\emph{Mid-left}: the kinetic temperatures derived 
from the H$_{2}$CO line ratios, represented by the color bar in units of Kelvin.
The blue solid line indicates the location of the assumed ionization front.
\emph{Mid-right}: maps of nonthermal line width (color bar in units of km\,s$^{-1}$).
\emph{Right}: map of Mach number. 
The black contours show the integrated intensity of H$_{2}$CO\,3$_{03}$--2$_{02}$ (same as in Fig.\,\ref{fig:H2CO-intensity-maps}).
}
\label{fig:H2CO-ratio}
\end{figure*}

\subsection{H$_{2}$CO line ratios}
\label{Sect:H2CO-Ratio}
As mentioned in Sect.\,\ref{Sect:Observations-and-data-reduction}, the line ratios 
H$_{2}$CO\,3$_{22}$--2$_{21}$/3$_{03}$--2$_{02}$ and 3$_{21}$--2$_{20}$/3$_{03}$--2$_{02}$ demonstrate similar 
patterns in determining the kinetic temperature of dense gas (e.g., \citealt{Mangum1993,Tang2017a}). In this 
work, we therefore utilize the average ratio of H$_{2}$CO\,0.5$\times$[(3$_{22}$--2$_{21}$\,+\,3$_{21}$--2$_{20})$/3$_{03}$--2$_{02}$] 
derived from the combined H$_{2}$CO\,3$_{22}$--2$_{21}$ and 3$_{21}$--2$_{20}$ to H$_{2}$CO\,3$_{03}$--2$_{02}$ 
velocity-integrated intensities as an indicator of the kinetic temperature. We selected the combined 
H$_{2}$CO\,3$_{22}$--2$_{21}$ and 3$_{21}$--2$_{20}$ lines, which were detected at S/Ns above 3$\sigma$. 
Generally, higher H$_{2}$CO line ratios indicate elevated kinetic temperatures \citep{Ao2013,Ginsburg2016,Tang2018a,Tang2021}. 
Consequently, the ratio maps can serve as a proxy for the relative kinetic temperature. 
The line ratio map of H$_{2}$CO in M17SW is shown in the top left panel of Fig.\,\ref{fig:H2CO-ratio}.
The line ratios range from 0.10 to 0.50 with an average value of 0.23\,$\pm$\,0.01.
Higher line ratios ($>$\,0.3) are found at the eastern edge of the primary structure of M17SW, the edge of 
clump G15.01-0.71, clumps G15.03-0.67 and G14.99-0.70, the southern regions of clumps G15.03-0.65, G15.02-0.68, 
G15.01-0.67, and G15.01-0.69, as well as the nearby ultra-compact H\,{\scriptsize II} region UC1 and the H$_2$O and 
CH$_3$OH masers. Moderate line ratios ranging from 0.2 to 0.3 are observed widely spread in the periphery of 
the primary structure of M17SW. Lower line ratios (<\,0.2) are found at the western edge of M17SW as well as the 
edges of clumps G15.01-0.71 and G14.99-0.70.

\subsection{Kinetic temperatures derived from H$_{2}$CO line ratios}
\label{Sect:Tkin}
As previously mentioned, the determination of the kinetic temperature involves calculating the averaged ratio of
H$_{2}$CO\,0.5$\times$[(3$_{22}$--2$_{21}$\,+\,3$_{21}$--2$_{20}$)/3$_{03}$--2$_{02}$]. To establish the relationship between the gas
kinetic temperature and the observed average of H$_{2}$CO\,0.5$\times$[(3$_{22}$--2$_{21}$\,+\,3$_{21}$--2$_{20}$)/3$_{03}$--2$_{02}$]
ratios, we employed the non local thermodynamic equilibrium (LTE) model RADEX\footnote{\tiny http://var.sron.nl/radex/radex.php}
\citep{Van2007} with collision rates obtained from \cite{Wiesenfeld2013}. In our analysis, we adopted a background 
temperature of 2.73\,K, an average measured line width of 3.8\,km\,s$^{-1}$, an H$_{2}$ number density of 
$n$(H$_{2}$)\,=\,5.5\,$\times$\,10$^{5}$\,cm$^{-3}$, and a para-H$_{2}$CO column density of 
$N$(para-H$_{2}$CO)\,=\,6.5\,$\times$\,10$^{13}$\,cm$^{-2}$ in Fig.\,\ref{fig:Radex}. 
The corresponding volume and column densities were derived from the following analysis.

We assumed that the para-H$_{2}$CO\,(3--2) lines are optically thin in the M17SW region. The average H$_{2}$ volume density and H$_{2}$CO
column density of the entire M17SW region were determined using multiple transitions of ortho-H$_{2}$CO\,($J$\,=\,2--1, 3--2, and 4--3) with
an angular resolution of $\sim$60$''$ \citep{Mundy1987}. The measured $n$(H$_{2}$) ranged from 10$^{4.3}$ to 10$^{6.4}$\,cm$^{-3}$ 
with an average of 5.5\,$\times$\,10$^{5}$\,cm$^{-3}$, which is consistent with results obtained using CS and C$^{18}$O 
\citep{Mundy1986,Stutzki1990,Wang1993}. We assumed that the para-H$_{2}$CO\,(3--2) lines observed originated from the same region 
with an average density of 5.5\,$\times$\,10$^{5}$\,cm$^{-3}$ in the M17SW region as measured by \cite{Mundy1987}. 
Furthermore, the derived $N$(ortho-H$_{2}$CO) ranges from 10$^{13.4}$ to 10$^{14.9}$\,cm$^{-2}$ with an average of 
1.94\,$\times$\,10$^{14}$\,cm$^{-2}$ in the M17SW region \citep{Mundy1987}. 
Previous observations of H$_{2}$CO\,($J$\,=\,3--2 and 4--3) with the APEX 12\,m telescope towards the dense clump G15.03-0.67 in M17SW
indicated $n$(H$_{2}$), $N$(para-H$_{2}$CO), and the ortho-to-para ratio of H$_{2}$CO to be 1.3\,$\times$\,10$^{6}$\,cm$^{-3}$,
6.2\,$\times$\,10$^{13}$\,cm$^{-2}$, and $\sim$3, respectively \citep{Tang2018b}. Based on these results, we assume the ortho-to-para
ratio of H$_{2}$CO to be 3 in the entire M17SW region. Therefore, we calculated an average $N$(para-H$_{2}$CO) of
6.5\,$\times$\,10$^{13}$\,cm$^{-2}$ in the M17SW region based on previous measurements by \cite{Mundy1987}. Using the methodology
outlined in Sect.\,3.1 of \cite{Tang2017b}, we employed RADEX to recalculate the average column density $N$(para-H$_{2}$CO)
of the entire M17SW region based on the H$_{2}$CO\,(3$_{03}$--2$_{02}$) average brightness temperature. The resulting value is
approximately 4$\times$10$^{13}$\,cm$^{-2}$ at $n$(H$_{2}$)\,=\,5.5\,$\times$\,10$^{5}$\,cm$^{-3}$, which agrees well with the results 
obtained from \cite{Mundy1987}. The H$_{2}$ column density derived from {\it Herschel} far infrared (FIR) continuum data
ranges from 1.0 to 4.0\,$\times$\,10$^{23}$\,cm$^{-2}$ with a median value of $\sim$2\,$\times$\,10$^{23}$\,cm$^{-2}$ in the M17SW region \citep{Hoang2022}.
Consequently, we calculated a typical fractional abundance of para-H$_{2}$CO to be approximately 3.3\,$\times$\,10$^{-10}$ in the M17SW
region based on the above average column density of para-H$_{2}$CO. This result aligns with the typical observational value of
$\sim$3.9\,$\times\,$10$^{-10}$ at various evolutionary stages of star formation regions \citep{Tang2018b}. 
Therefore, the adopted average $N$(para-H$_{2}$CO) of the entire M17SW region is deemed reasonable in this study.

Previous observations regarding kinetic temperature have revealed subtle differences between various column densities at a 
given $n$(H$_{2}$) \citep{Tang2018b}. In this study, we examine the relationship between kinetic temperature and the line ratios
of para-H$_{2}$CO at $N$(para-H$_{2}$CO) values of 1.0, 6.5, and 13.0\,$\times$\,10$^{13}$\,cm$^{-2}$, all at an $n$(H$_{2}$) of 
5.5\,$\times$\,10$^{5}$\,cm$^{-3}$, as illustrated in Fig.\,\ref{fig:Radex}. These comparisons highlight nuanced temperature 
variations across different column densities. The temperature differences derived from column densities of 
1.0\,$\times$\,10$^{13}$ and 1.3\,$\times$\,10$^{14}$\,cm$^{-2}$, in comparison to 6.5\,$\times$\,10$^{13}$\,cm$^{-2}$, 
are found to be smaller than the uncertainties associated with temperature estimations based on line ratio measurements. 
Given the potential variability in the relationship between gas temperature and the H$_{2}$CO line ratio across different spatial densities, as illustrated in Fig.\,4 of \cite{Tang2018b}, we conducted modeling at spatial densities of $2\times10^4$ and $2.5\times10^6$\,cm$^{-3}$ based on previously measured $n$(H$_{2}$) values \citep{Mundy1987} using the RADEX. It appears that the $T_{\rm kin}$ measured at $5.5\times10^5$\,cm$^{-3}$ as utilized in this study exhibits a slight variation of less than 15\% at spatial densities of $2\times10^4$ and $2.5\times10^6$\,cm$^{-3}$ for $T_{\rm kin}$ values below 100\,K.
Consequently, we could transform the line ratio map into a kinetic temperature map using a column density of 
6.5\,$\times$\,10$^{13}$\,cm$^{-2}$ and an H$_{2}$ number density of 5.5\,$\times$\,10$^{5}$\,cm$^{-3}$. 
The maps showing the H$_{2}$CO line ratios and the corresponding kinetic temperatures derived from the H$_{2}$CO line ratios
are presented in Fig.\,\ref{fig:H2CO-ratio}.

The validity of LTE as a dependable approximation for H$_{2}$CO level populations has been established under specific conditions, 
particularly when the relevant transitions are optically thin and the systems exhibit high densities \citep{Mangum1993}. 
Following the methodology described in \cite{Tang2017b} (refer to their Eq.\,(2)) and utilizing the RADEX model, 
we conducted a thorough analysis to explore the correlation between the kinetic temperature and the combined H$_{2}$CO 
line ratio within the LTE regime, as illustrated in Fig.\,\ref{fig:Radex}. It reveals a satisfactory agreement between 
the temperatures derived from the LTE model and those obtained from the RADEX model, assuming a column density of 
6.5\,$\times$\,10$^{13}$\,cm$^{-2}$, provided that $T_\mathrm{kin}$ remains below approximately 200\,K.

The kinetic temperature in M17SW, as indicated by the H$_{2}$CO line ratios, ranges from 27 to 181\,K, with an average 
value of 54.2\,$\pm$\,0.3\,K. The kinetic temperature map in Fig.\,\ref{fig:H2CO-ratio} reveals that the highest 
temperature ($T_\mathrm{kin}$>100\,K) is associated with UC1 and the eastern edge of M17SW. 
High temperatures (>70\,K) are detected at the eastern edge of the primary structure of M17SW, the edge of 
clump G15.01-0.71, clumps G15.03-0.67 and G14.99-0.70, the southern regions of clumps G15.03-0.65, G15.02-0.68, G15.01-0.67, 
and G15.01-0.69, as well as near H$_2$O and CH$_3$OH masers. Moderate temperatures ranging from 50 to 70\,K 
are widely observed in the periphery of the primary structure of M17SW. Lower temperatures (<50\,K) are found 
at the western edge of M17SW and the edges of clumps G15.01-0.71 and G14.99-0.70.

Previous observations indicate that the H$_{2}$CO\,3$_{03}$--2$_{02}$ emission may be optically thick in the dense region
(e.g., \citealt{Immer2016,Tang2018b}). This saturation could result in falsely high H$_{2}$CO\,3$_{22}$--2$_{21}$/3$_{03}$--2$_{02}$ and 3$_{21}$--2$_{20}$/3$_{03}$--2$_{02}$ line ratios in our integrated intensity ratio map, potentially mimicking elevated gas temperatures in the clouds. 
As stated by \cite{Yue2026}, the H$_{2}$CO\,(3--2) lines may overestimate the kinetic temperature when $T_{\rm kin}$\,$>$\,100\,K.
To address this issue, we utilized the RADEX non-LTE model to calculate the opacity of H$_{2}$CO\,3$_{03}$--2$_{02}$ line, adopting an average temperature of $T_{\rm kin}$\,=\,54\,K, an average measured linewidth of 3.8\,km\,s$^{-1}$, and a gas volume density of $n$(H$_2$)\,=\,5.5\,$\times$\,10$^{5}$\,cm$^{-3}$. The para-H$_{2}$CO column density was taken as $N$(para-H$_{2}$CO)\,=\,8.4$\times$10$^{12}$--2.6$\times$10$^{14}$\,cm$^{-2}$, derived from previous ortho-H$_{2}$CO measurements \citep{Mundy1986} using an ortho-to-para ratio of 3 \citep{Tang2018b}. The resulting optical depths for the H$_{2}$CO\,3$_{03}$--2$_{02}$ line range from approximately 0.08 to 1.5. These values suggest that in the dense cores, 
opacities in the H$_{2}$CO\,3$_{03}$--2$_{02}$ lines may exhibit slight saturation, potentially leading to a minor overestimation of kinetic temperatures by less than 20\%.

\subsection{Thermal and non-thermal motions obtained from H$_{2}$CO}
\label{Sect:T&NT-thermal}
The thermal and non-thermal line widths \citep{Pan2009,Dewangan2016} can be determined by the kinetic temperature
obtained from the line ratios of H$_{2}$CO. We used $\sigma_{\rm T}$\,=\,$\sqrt{\frac{kT_{\rm kin}}{m_{\rm H_2CO}}}$ and 
$\sigma_{\rm NT}$\,=\,$\sqrt{\frac{\Delta v^2}{8{\rm ln}2}-\sigma_{\rm T}^2}$ to calculate the thermal and non-thermal 
line widths in M17SW, respectively. In these formulas,  $k$ represents the Boltzmann constant, $T_{\rm kin}$ stands 
for the kinetic temperature of the gas, $m_{\rm H_2CO}$ denotes the mass of a formaldehyde molecule, and $\Delta v$ 
indicates the measured FWHM line width of H$_{2}$CO\,3$_{03}$--2$_{02}$. 
As outlined in Sect.\,\ref{Sect:Observations-and-data-reduction},
there are roughly 330 pixels (accounting for roughly 26\%) exhibiting two velocity components at 18.5 and 21.5\,km\,s$^{-1}$,
aligning with previous findings from NH$_3$ and CO studies \citep{Guesten1988,Stutzki1990,Keown2019}. 
Due to the proximity and low intensity of these two velocity components, 
neither the H$_{2}$CO 3$_{22}$--2$_{21}$ and 3$_{21}$--2$_{20}$ nor the combined 3$_{22}$--2$_{21}$ and 3$_{21}$--2$_{20}$ 
lines can effectively differentiate between the two velocity components.
The precise determination of kinetic temperatures in regions characterized by dual velocity components presents challenges.
Therefore, the line width is obtained as a weighted average of the line widths of the two velocity components within 
H$_{2}$CO\,3$_{03}$--2$_{02}$ based on their integrated intensities.
The thermal line width ranges from 0.09 to 0.22\,km\,s$^{-1}$, with an average value of 0.12\,$\pm$\,0.01\,km\,s$^{-1}$.
Simultaneously, the non-thermal line width ranges from 0.87 to 3.2\,km\,s$^{-1}$, with an average value of 1.67\,$\pm$\,0.01\,km\,s$^{-1}$.
It is important to notice that the non-thermal line width surpasses the thermal line width.
These results are consistent with previous H$_{2}$CO ($J$=3--2 and 4--3) observations of the massive star-forming 
regions in our Galaxy and Large Magellanic Cloud \citep{Tang2017b,Tang2018a,Tang2018b,Tang2021,Zhao2024}.

The non-thermal line width distribution map of H$_{2}$CO\,3$_{03}$--2$_{02}$ is depicted in Fig.\,\ref{fig:H2CO-ratio}. 
This distribution is akin to the line width results observed in M17SW, with the exception of regions near the dense 
clumps G15.03-0.65, G15.01-067, and G15.02-0.68. As mentioned before, there are two velocity components in the central 
area of M17SW. In these specific locations, the non-thermal line width displays lower values, possibly due to the weighted
average of the line widths of the two velocity components. 
The systematically lower $\sigma_{\rm NT}$ derived from two-component fits arises because the single-component alternative, 
when forced to fit a blended profile with a weak secondary peak, 
overestimates the line width by conflating the intrinsic turbulence with the unresolved velocity separation between components.
Conversely, the non-thermal line width demonstrates higher values 
near clump G15.03-0.67, G15.03-0.65, G14.99-0.67, and G14.99-0.70, as well as along the northern and western edges of M17SW. This discrepancy could be attributed to the relatively lower S/Ns of
the spectral lines, rendering Gaussian profile fitting more challenging.

The Mach number ($\mathcal{M}$\,=\,$\sigma_{\rm NT}/c_{\rm s}$, in which 
$c_{\rm s}$\,=\,$\sqrt{\frac{kT_{\rm kin}}{\mu m_{\rm H}}}$, 
where $\mu$\,=\,2.37 is the mean molecular weight for molecular clouds and $m_{\rm H}$ is the mass of a hydrogen atom) 
distribution map of M17SW is shown in Fig.\,\ref{fig:H2CO-ratio}. The Mach number ranges from 1.3 to 8.5 with an average of 4.0$\pm$0.1.
This indicates that the dense gas in M17SW traced by H$_{2}$CO is dominated by supersonic non-thermal motions 
(e.g., turbulence, outflows, shocks). High Mach numbers ($\mathcal{M}$\,>\,6) are observed in 
the vicinity of the clumps G15.03-0.67, G15.03-0.65, G14.99-0.70, G14.99-0.67, and G14.98-0.69, as well as along the northern and western edges of M17SW. Medium Mach numbers 
(4\,<\,$\mathcal{M}$\,<\,6) are found to be associated with the dense clumps. Low Mach numbers ($\mathcal{M}$\,<\,4) 
are prevalent in the dense region of the primary structure and the eastern and southeastern edge of M17SW.

\section{Discussion}
\label{Sect:Discussion}
\subsection{Comparison of temperatures derived from gas and dust}
\label{Sect:Tk-H2CO-others}
As delineated in Appendix\,\ref{Sect:Previous} and Table\,\ref{table:M17SW-previous-T}, \cite{Guesten1988} employed 
the Effelsberg 100\,m telescope with 
a beam size of $\sim$\,40$''$ to conduct a temperature structure mapping of the M17SW region using multiple transitions
of NH$_{3}$\,(1,1)--(6,6). Their findings suggest that temperatures within the M17SW molecular cloud vary between 
approximately 30 and 275\,K, with a bulk temperature of around 50\,K. 
Approximately 10\% of the gas populates the NH$_{3}$\,(6,6) inversion transition level, which leads to a notably elevated measured temperature.
Our temperature assessments derived from 
H$_{2}$CO\,(3--2) line ratios align closely with their reported outcomes. More recently, the temperature distributions 
of the M17SW region were also observed using NH$_{3}$\,(1,1) and (2,2) with the GBT 
(beam size $\sim$32$''$, $\sim$0.32\,pc; \citealt{Keown2019}). 
The temperature map derived from the NH$_{3}$\,(2,2)/(1,1) line ratios is depicted in Fig.\,\ref{fig:NH3_T} 
(also see Fig.\,11 in \citealt{Keown2019}). The kinetic temperature of M17SW as determined from NH$_{3}$\,(2,2)/(1,1) observations 
ranges from 20 to 80\,K with an average of $\sim$40\,K. Notably, temperatures exceeding 50\,K are observed around the 
clumps G15.03-0.67, G15.03-0.65, and G15.01-0.71, while temperatures ranging from 30 to 40\,K are prevalent around the 
clumps G15.02-0.68, G15.01-0.67, and G15.01-0.69. Lower temperatures of $\sim$20\,K are identified in the clumps G14.99-0.67,
G14.98-0.69, and G14.99-0.70. The temperature gradient derived from NH$_{3}$ in M17SW shows a decrease from east (>70\,K) 
to west ($\sim$20\,K), displaying a similar distribution compared to the temperature distribution obtained from H$_{2}$CO\,(3--2) 
line ratios on a large scale. As mentioned in Sect.\,\ref{Sect:Tkin}, the kinetic temperature derived from H$_{2}$CO 
near the ultra-compact H\,{\scriptsize II} region UC1 reaches >100\,K, and it decreases to about 30--40\,K at the western edge 
of M17SW. The average temperature derived from NH$_{3}$\,(2,2)/(1,1) line ratios of the M17SW region is slightly lower 
than that from the H$_{2}$CO\,(3--2) line ratios. The high kinetic temperatures obtained from H$_{2}$CO are associated with 
UC1, dense clumps, as well as H$_2$O and CH$_3$OH masers (see Sect.\,\ref{Sect:Tkin}), displaying a similar distribution 
to NH$_{3}$. For the dense clumps G15.02-0.68, G15.01-0.67, G15.01-0.69, G14.98-0.69, and G14.99-0.70, the kinetic temperatures obtained from H$_{2}$CO 
show a different distribution compared to NH$_{3}$, all exhibiting significantly elevated temperatures from H$_{2}$CO relative 
to NH$_{3}$\,(2,2)/(1,1). These observed differences may arise from the varying beam sizes of the two observations
and/or the fact that higher-excitation H$_{2}$CO lines are more sensitive to warmer gas components.
The energy levels of the H$_{2}$CO\,3$_{03}$--2$_{02}$, 3$_{22}$--2$_{21}$, and 3$_{21}$--2$_{20}$ lines are 21, 68, and 68\,K, respectively, whereas NH$_3$\,(1,1) and (2,2) exhibit $E_{\rm u}$ of 1 and 42\,K. The elevated excitation energy of H$_2$CO renders it a more specific tracer of warm dense gas, as indicated by \cite{Ginsburg2016}.

\begin{figure*}[t]
\centering
\includegraphics[width=0.95\textwidth]{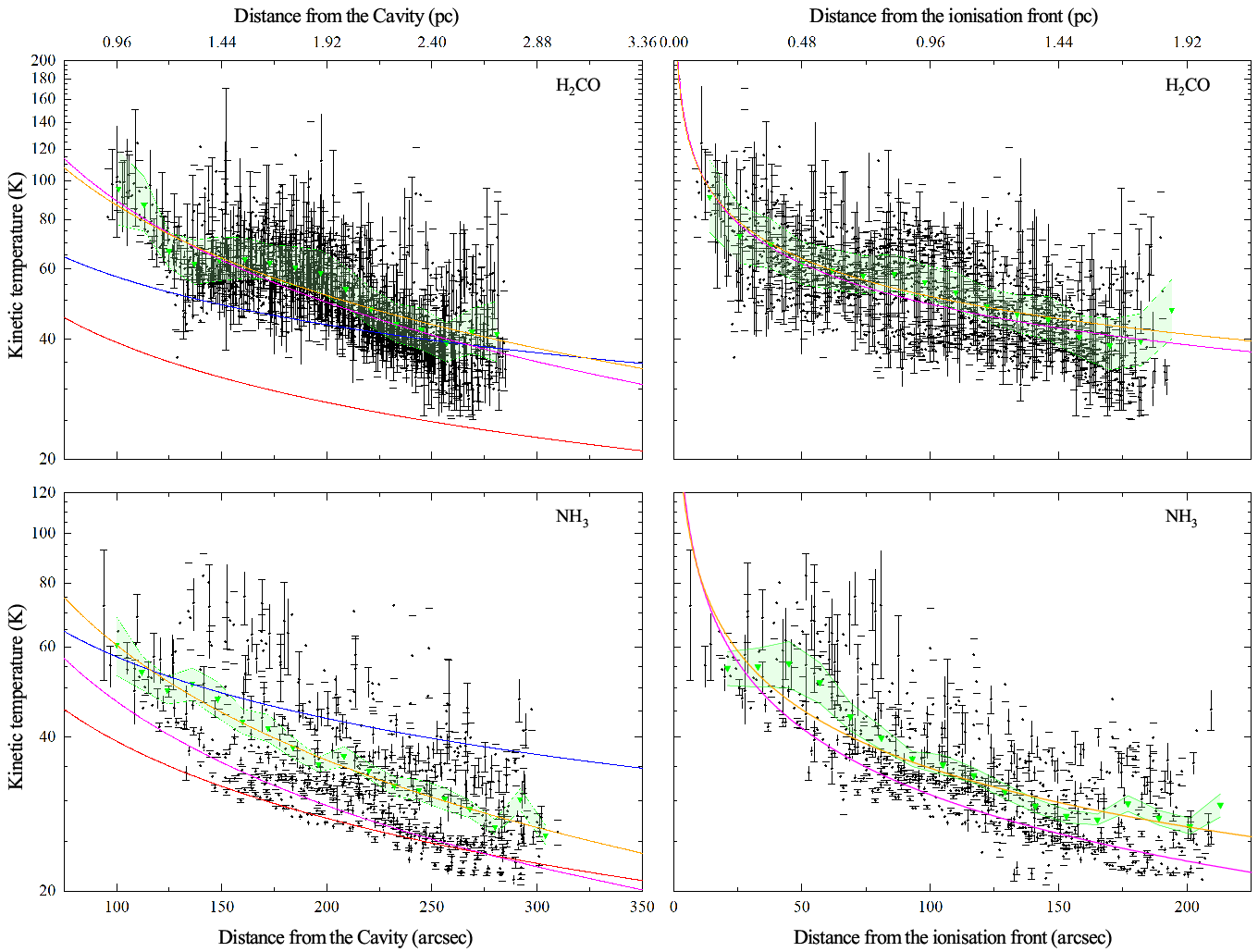}
\caption{Variation in kinetic temperature across the M17SW region with distance from the NGC\,6618 cavity (\emph{left panels}) and the assumed ionization front (\emph{right panels}) (see Sect.\,\ref{Sect:Radiative-heating}). The kinetic temperatures are calculated from the line ratios of H$_{2}$CO\,(3--2) (\emph{upper panels}) and NH$_{3}$\,(2,2)/(1,1)
(\emph{lower panels}) (for details, see Sect.\,\ref{Sect:Radiative-heating}). The average kinetic temperature at each projected distance is denoted by green triangles within 12$''$ bins ($\sim$0.12\,pc), plotted at the bin centers, with the green shaded area representing the corresponding uncertainty. The magenta and orange lines represent the fitted results for all data and binned data, respectively, with the regression
equation formatted as $T_{\rm kin}\,=\,a\times(\frac{R}{\rm arcsec})^{b}~{\rm K}$. The fitting method employed is the Levenberg-Marquardt algorithm, which accounts for temperature errors through instrumental weighting ($w = 1/\sigma^2$). 
The reduced Chi-squared values for all H$_{2}$CO and NH$_{3}$ data obtained from the cavity are 4.14 and 33.38, respectively. From the ionization front, these values are 5.79 and 25.21 for H$_{2}$CO and NH$_{3}$, respectively. For the binned data, the reduced Chi-squared values for H$_{2}$CO and NH$_{3}$ from the cavity are 0.34 and 0.73, while from the ionization front, they are 0.22 and 1.97, respectively. The red and blue lines depict the expected relationships derived from 
the Stefan-Boltzmann law and the modified Stefan-Boltzmann law, respectively. 
}
\label{fig:cavity-distance}
\end{figure*}

The CO observations from highly excited rotational levels (i.e., $J$\,=\,7--6 and 14--13) reveal the presence of warm and dense gas in various regions of 
M17SW \citep{Jaffe1987,Harris1987}, with a complex molecular cloud structure surrounding the interface \citep{Stutzki1988}. 
Analysis of CO\,($J$=7--6 and 14--13) and [OI] data indicates a kinetic temperature exceeding 200\,K at the interface 
between the M17SW cloud and the OB star cluster \citep{Harris1987,Stutzki1988}, consistent with results obtained 
from H$_{2}$CO line ratios near UC1. Modeling of $^{13}$CO\,$J$\,=\,1--0 and 
5--4 emissions reveals that the eastern edge of M17SW exhibits the highest gas temperature around 60\,K, 
decreasing to $\sim$20\,K at the western edge \citep{Howe2000}. These temperatures, while relatively lower 
compared to those determined by H$_{2}$CO, are similar to those from NH$_{3}$\,(2,2)/(1,1) measurements, 
all showing a consistent temperature 
gradient with H$_{2}$CO and NH$_{3}$. This suggests that the heating of the M17SW molecular cloud is influenced 
by external sources, particularly in the northeastern region where the OB star cluster is located.

As mentioned in Appendix\,\ref{Sect:Previous} and Table.\,\ref{table:M17SW-previous-T}, 
the dust temperature ($T_\mathrm{\rm dust}$) in the M17SW region has been investigated across 
various wavelengths ranging from 50 to 850\,$\mu$m (e.g., \citealt{Gatley1979,Meixner1992,Dupac2002,Hoang2022}). 
These studies reveal that the typical dust temperature in the M17SW region falls within the range of approximately 
20--120\,K, akin to the gas temperatures derived from NH$_{3}$\,(2,2)/(1,1) \citep{Guesten1988,Keown2019} and 
low--$J$ $^{13}$CO \citep{Howe2000}, albeit slightly lower than the values obtained from 
high-$J$ CO \citep{Harris1987,Stutzki1988} and H$_{2}$CO line ratios. This consistency aligns with previous
findings in star-forming regions within our galaxy (e.g., \citealt{Tang2017b,Tang2018a,Tang2018b,Zhao2024}). 
The dust temperature decreases from east to west within the M17SW cloud, with the highest dust temperature 
observed at the interface between the M17SW cloud and the OB stars 
(e.g., \citealt{Harris1987,Guesten1988,Stutzki1988,Meixner1992,Dupac2002,Hoang2022}). This pattern mirrors 
the kinetic temperature distribution inferred from H$_{2}$CO line ratios and NH$_{3}$ on a larger scale, 
indicating a consistent heating effect from massive star clusters that is prominently visible in large-scale features.

\begin{figure*}[t]
\centering
\includegraphics[width=0.95\textwidth]{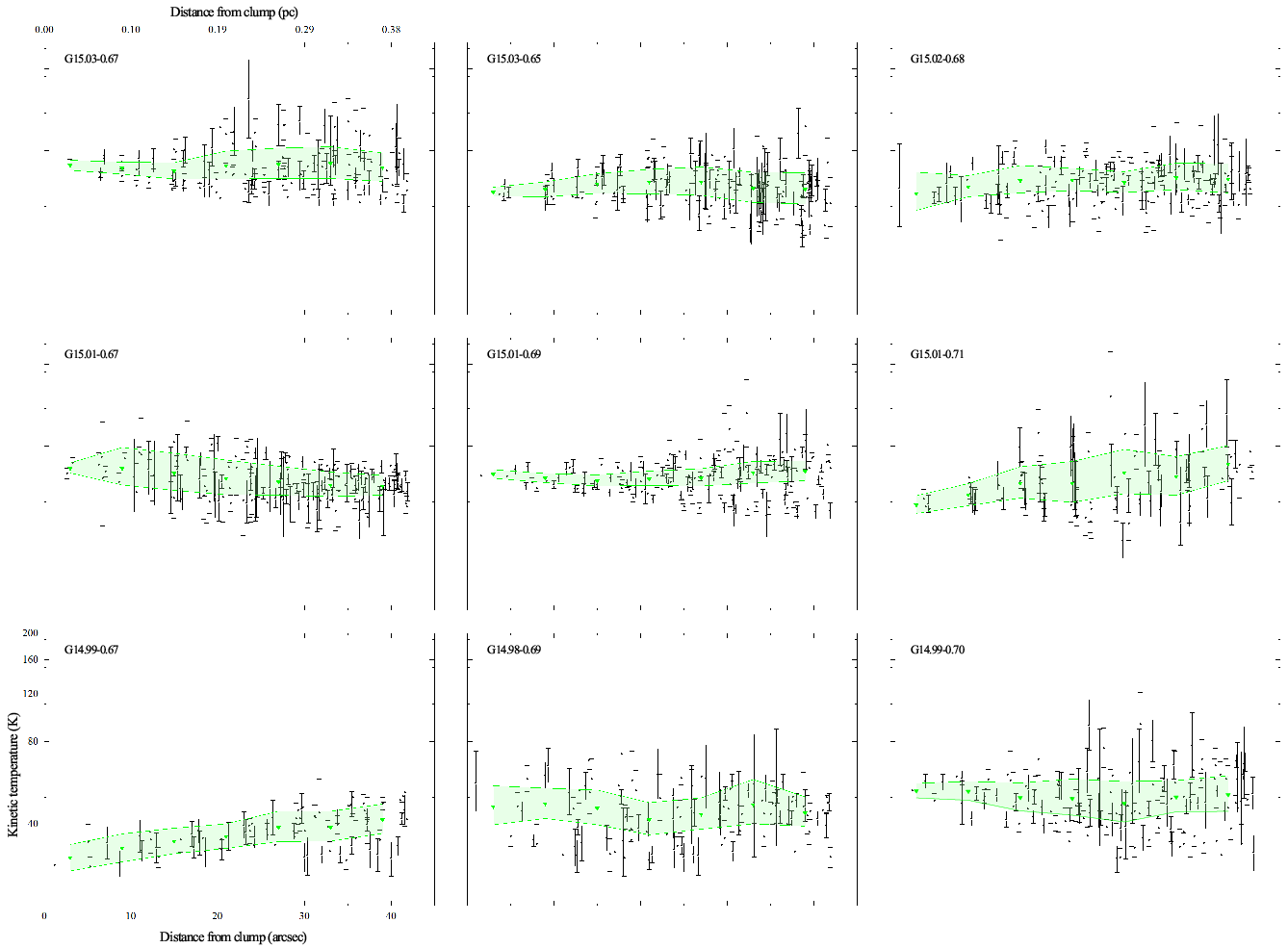}
\caption{Variations of the gas kinetic temperatures derived from H$_{2}$CO\,(3--2) line ratios,
measured from the central regions of dense clumps (see Table\,\ref{table:Clump-Parameters}) extending towards their respective peripheries.
The green downward-pointing triangles represent the mean temperatures calculated in bins of 6$''$, with the green shaded areas indicating their uncertainties. 
The distances are plotted at the bin centers. To prevent cross-contamination from adjacent clumps, the data were restricted to the region on each clump's side of the midline between the two clumps.
}
\label{fig:h2co_clump}
\end{figure*}

\subsection{Radiative heating}
\label{Sect:Radiative-heating}

As discussed above, the M17SW molecular cloud is influenced by UV radiation emitted from OB star clusters 
in the northeast, leading to a temperature gradient that decreases from northeast to southwest within the cloud. 
To investigate the impact of external radiation on the heating of the M17SW molecular cloud, we analyzed the 
temperature profiles of H$_{2}$CO and NH$_{3}$ (see Fig.\,\ref{fig:cavity-distance}). The location of the "cavity" 
is considered as the central point of the main source of the radiation source within the NGC\,6618 open cluster (also see Fig.\,13 in \citealt{Lim2020}), 
surrounded by the M17 H\,{\scriptsize II} region. This cavity, characterized by hot and diffuse mid-infrared 
and sub-millimeter emissions, as well as significant molecular line emission, is centered at $\alpha_{2000}$\,=\,18$^h$20$^m$31\hbox{$\,.\!\!^s$}2
and $\delta_{2000}$\,=\,-16$\degr$10$'$54\hbox{$\,.\!\!^{\prime\prime}$}8 \citep{Townsley2003,Povich2007,Lim2020}.
Gas temperatures were obtained from measurements of H$_{2}$CO\,(3--2) and NH$_{3}$\,(2,2)/(1,1) line ratios \citep{Keown2019}. 
For H$_{2}$CO data, the temperature uncertainties do not exceed 50\% of the temperature value. Regarding NH$_{3}$ data, 
the S/Ns of NH$_{3}$\,(2,2) are $\gtrsim$\,4 and the temperature errors are within 30\% of the temperature value. 
Due to the limited reliability of high kinetic temperatures indicated by NH$_{3}$\,(2,2)/(1,1) line ratios 
(e.g., \citealt{Tang2017b,Tang2018a,Tang2018b,Zhao2024}), only NH$_{3}$ data with temperatures below 100\,K were selected.
As depicted in Fig.\,\ref{fig:cavity-distance}, the kinetic temperatures probed by H$_{2}$CO and NH$_{3}$ 
decrease noticeably with increasing projected distance ($R$) from the cavity. A power-law function is adopted here as an empirical description of the observed temperature decrease with distance. 
This form is scale-free, meaning it does not assume any characteristic length scale, and allows a direct comparison with the 
theoretical curves derived from the Stefan-Boltzmann blackbody radiation law. Such a function has been widely applied to model 
temperature gradients in massive star-forming regions 
(e.g., \citealt{Busquet2016,vantHoff2018,Tang2018a,Tobin2020,Gieser2021,Gieser2022,Gieser2023,Zhao2024}).
The fitted results of H$_{2}$CO and NH$_{3}$ are
\begin{eqnarray}
\label{equation:H2CO}
T_{\rm kin} ({\rm H_{2}CO}) = 4413(\pm487) \times \bigg(\frac{R}{\rm arcsec}\bigg)^{-0.85\pm0.02}~{\rm K}, 
\end{eqnarray}
\begin{eqnarray}
\label{equation:NH3}
T_{\rm kin} ({\rm NH_{3}}) = 1057(\pm119) \times \bigg(\frac{R}{\rm arcsec}\bigg)^{-0.68\pm0.02}~{\rm K},
\end{eqnarray}
with power-law indices of --0.85 and --0.68, respectively.
The gas temperatures derived from H$_{2}$CO and NH$_{3}$ at projected distance are averaged within 12$''$ bins ($\sim$0.12\,pc) as shown in Fig.\,\ref{fig:cavity-distance}. Additionally, the average gas temperature data was fitted, 
yielding fitted results of
\begin{eqnarray}
\label{equation:H2CO_bin}
T_{\rm kin} ({\rm H_{2}CO}) = 2788(\pm1040) \times \bigg(\frac{R}{\rm arcsec}\bigg)^{-0.75\pm0.07}~{\rm K}, 
\end{eqnarray}
\begin{eqnarray}
\label{equation:NH3_bin}
T_{\rm kin} ({\rm NH_{3}}) = 1886(\pm370) \times \bigg(\frac{R}{\rm arcsec}\bigg)^{-0.75\pm0.04}~{\rm K},
\end{eqnarray}
with the same power-law index of --0.75.
These power-law indices are consistent with previous observations in various star formation regions,
which typically range from -0.05 to -1.48 (e.g., \citealt{Tang2018a,Gieser2019,Gieser2021,Gieser2022,Gieser2023,Lin2022,Zhao2024}),
directly indicating that the dense gas in the PDR M17SW is heated by 
radiation emitted from OB star 
clusters in the cavity. These relations are based on the projected separation from the radiation source.

As the distance changes, the expected variation in gas temperature is predicted to adhere to the principles of the
Stefan-Boltzmann blackbody radiation law ($T_{\rm kin}=0.86\times(\frac{L}{{\rm L}_{\odot}})^{1/4}(\frac{R}{\rm pc})^{-1/2}~{\rm K}$, 
where the luminosity $L$ is given in terms of solar luminosity (${\rm L}_{\odot}$) and the distance $R$ is measured in parsecs). 
By refining the emissivity of dust grains to wavelengths below the characteristic black-body temperature, the radiation 
law can be modified to $T_{\rm kin}=2.7\times(\frac{L}{{\rm L}_{\odot}})^{1/5}(\frac{R}{\rm pc})^{-2/5}~{\rm K}$
\citep{Wiseman1998,Tang2018a}.
Considering the total luminosity estimated ranging from 4.5 to 5.6\,$\times$\,10$^{6}\,\mathrm{L}_{\odot}$
at a distance of 2.2\,kpc \citep{Povich2007}, we adopt a representative value of 
4$\,\times\,$10$^{6}$\,${\rm L}_{\odot}$ for the M17 region at a distance of 2\,kpc in this study. Our fitted temperature
result from H$_{2}$CO line ratios aligns with the modified Stefan-Boltzmann blackbody radiation law 
(see Fig.\,\ref{fig:cavity-distance}). However, the NH$_{3}$ 
data show better agreement with the Stefan-Boltzmann blackbody radiation law. Comparable results for H$_{2}$CO and NH$_{3}$ 
data were also observed in the OMC-1 region \citep{Tang2018a}. Similarly, the NH$_{3}$ data indicate two temperature components, 
"cold" and "warm," with distance from the cavity (see Fig.\,\ref{fig:cavity-distance}). The warm component corresponds
to the results derived from H$_{2}$CO concerning distance, which closely matches the predicted outcomes from the modified
Stefan-Boltzmann blackbody radiation law. Conversely, the cold component aligns with the predicted results from 
the unmodified Stefan-Boltzmann blackbody radiation law. 
Previous NH$_3$ observations have indicated the existence of both cold and warm gas components in M17SW  \citep{Guesten1988}. 
However, due to significant uncertainties in NH$_3$ data, a reliable and physically motivated separation of these components is not achievable.
As a result, we retain a single-component analysis for deriving the temperature gradient from NH$_3$ data.
Both radiation models for gas heating (Stefan-Boltzmann blackbody radiation 
and its modification related to dust emissivity) are well supported by our H$_{2}$CO and NH$_{3}$ data, making it challenging 
to discern which model is better.

In regions with distances $R$\,<\,125$''$ ($\sim$1.25\,pc) from the cavity, several locations exhibit gas temperatures probed with 
H$_{2}$CO and NH$_{3}$ above the fitted results (see Fig.\,\ref{fig:cavity-distance}). Currently, extensive shocked gas 
surrounding the M17 H\,{\scriptsize II} region is detected on a large scale, likely originating from the collision between 
the expanding ionized gas and the ambient neutral medium \citep{Zhu2023}. The high temperatures of the dense gas near the M17 
H\,{\scriptsize II} region may be attributed to shocks from the collisional interface and stellar winds from the H\,{\scriptsize II} 
region. For regions with $R$\,=\,160--240$''$ ($\sim$1.6--2.4\,pc), some locations display gas temperatures probed with H$_{2}$CO and NH$_{3}$ 
above the fitted results. These elevated temperatures are linked to dense gas structures in M17SW, particularly near 
dense clumps and H$_2$O and CH$_3$OH masers, suggesting that the dense gas may be heated by local star formation feedback 
such as radiation, outflows, shocks, and winds.

In addition to the distance from the cavity, we also examine the temperature gradient with respect to the assumed ionization front
(see Fig.\,\ref{fig:H2CO-ratio}). Based on the morphology of the cloud and the spatial distribution of the UV radiation field, the ionization front is defined along the northeastern rim of M17SW. The orientation of the assumed ionization front was determined by performing principal component analysis on the spatial coordinates of the points along the southeastern edge of M17SW. The first principal component gives the direction of the edge, which we adopt as the orientation of the ionization front. The projected distance of each pixel to this front is then calculated.
Fig.\,\ref{fig:cavity-distance} shows the kinetic temperature as a function of distance from the assumed ionization front. A decreasing trend is clearly visible, with higher temperatures near the front and lower temperatures farther away.

\begin{figure*}[!t]
\centering
\begin{minipage}[c]{0.7\textwidth}
  \centering
  \includegraphics[width=0.95\textwidth]{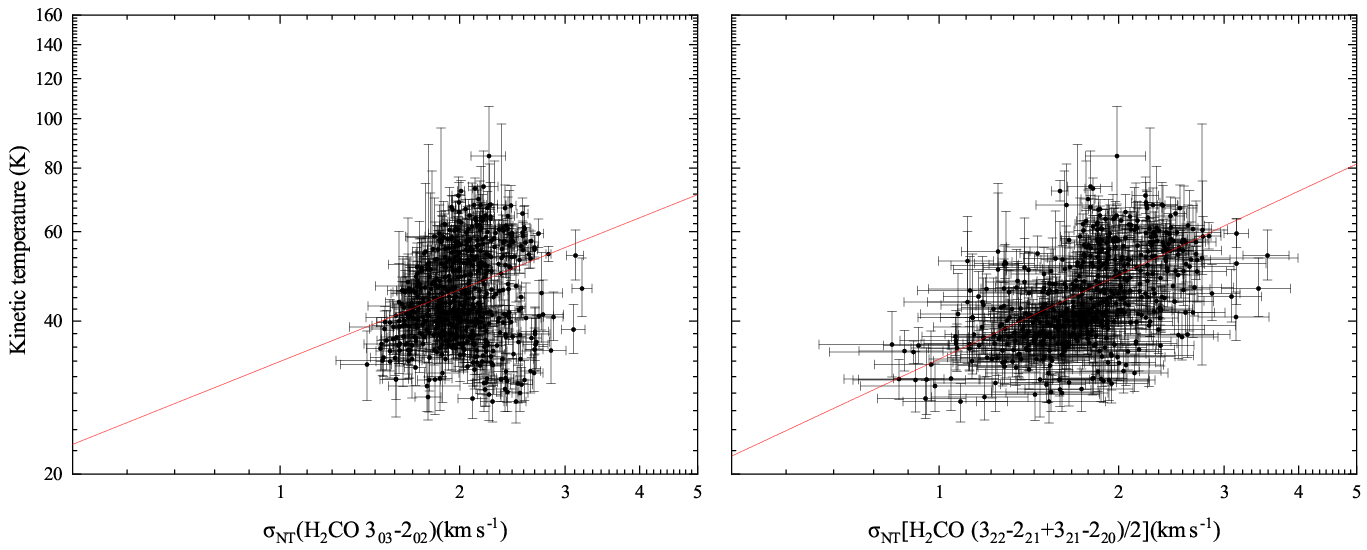}
\end{minipage}%
\hfill
\begin{minipage}[c]{0.30\textwidth}
  \caption{The correlation between the nonthermal line width ($\sigma_{\rm NT}$) and gas kinetic temperature ($T_{\rm kin}$) obtained from the H$_{2}$CO line ratio is investigated in the M17SW region characterized by Mach numbers $\mathcal{M}$\,$\gtrsim$\,4.0 (see Sect.\,\ref{Sect:Turbulent-heating}). A linear regression analysis with weighting ($w = 1/\sigma^2$) is conducted on the H$_{2}$CO\,3$_{03}$--2$_{02}$ (\emph{left})
and the combined H$_{2}$CO\,3$_{22}$--2$_{21}$ and 3$_{21}$--2$_{20}$ (\emph{right}) lines, resulting in: log$T_{\rm kin}=(0.47\pm0.06)\times{\rm log}\sigma_{\rm NT}+(1.52\pm0.02)$ and 
log$T_{\rm kin}=(0.55\pm0.04)\times{\rm log}\sigma_{\rm NT}+(1.53\pm0.01)$, respectively. The Pearson correlation coefficients ($r$) for these regressions (red lines) are 0.30 and 0.52, respectively.}
  \label{fig:NT-T}
\end{minipage}
\end{figure*}

We still perform power-law fits to the temperature–distance relations for both H$_2$CO and NH$_3$ to facilitate a qualitative comparison between the two tracers:
\begin{eqnarray}
\label{equation:H2CO_IF}
T_{\rm kin} ({\rm H_2CO}) = 234(\pm13) \times \left(\frac{R}{\rm arcsec}\right)^{-0.34\pm0.01}~{\rm K},
\end{eqnarray}
\begin{eqnarray}
\label{equation:NH3_IF}
T_{\rm kin} ({\rm NH_3}) = 225(\pm 11) \times \left(\frac{R}{\rm arcsec}\right)^{-0.43\pm 0.01}~{\rm K},
\end{eqnarray}
with power-law indices of –0.34 and –0.43, respectively. 
We also averaged the gas temperatures obtained from H$_2$CO and NH$_3$ at projected distance within 12$''$ bins in Fig.\,\ref{fig:cavity-distance}.
For the average gas temperature data, the fitted results are
\begin{eqnarray}
\label{equation:H2CO_IF_bin}
T_{\rm kin} ({\rm H_2CO}) = 220(\pm28) \times \left(\frac{R}{\rm arcsec}\right)^{-0.32\pm0.03}~{\rm K},
\end{eqnarray}
\begin{eqnarray}
\label{equation:NH3_IF_bin}
T_{\rm kin} ({\rm NH_3}) = 200(\pm 31) \times \left(\frac{R}{\rm arcsec}\right)^{-0.38\pm 0.03}~{\rm K},
\end{eqnarray}
with power-law indices of –0.32 and –0.38, respectively. 
These results are consistent with the scenario of external UV heating dominating the thermal balance in the M17SW region.
One should note that the ionization front is defined from the cloud morphology rather than being directly observed, and projection effects are non-negligible. Accordingly, the fitted temperature gradients should be interpreted with caution. Nevertheless, these fitted results provide a meaningful comparison between H$_2$CO and NH$_3$, and the derived gradients are broadly consistent with predictions from the modified Stefan–Boltzmann blackbody radiation law.

To investigate the impact of internal radiation heating on the dense gas surrounding dense clumps within the M17SW region at a scale of 0.1--0.4\,pc, we analyzed the temperature profiles of the nine dense clumps, already addressed in Table\,\ref{table:Clump-Parameters}, in Fig.\,\ref{fig:h2co_clump}. 
Among these nine clumps, six (G15.03-0.67, G15.03-0.65, G15.02-0.68, G15.01-0.67, G14.98-0.69, and G14.99-0.70) are associated with
maser emission (see Fig.\,\ref{fig:H2CO-intensity-maps}), indicating ongoing star formation activities.
G15.01-0.67 and G15.01-0.69 display weak temperature gradients that decrease with distance from the clump, suggesting that internal star formation activities may be responsible for heating the dense gas in these regions. In contrast, other dense clumps show no significant temperature gradient (that decrease with distance from the clump), possibly indicating that they are in an earlier evolutionary stage (such as Class\,I CH$_3$OH masers; \citealt{Kurtz2004,Gomez2010,Kim2018a,Yang2020}) with embedded protostars or clusters that have not yet generated sufficient heat to warm the surrounding molecular envelope significantly.
The bolometric luminosities of four dense clumps, namely G15.03-0.67, G15.03-0.65, G15.01-0.67, and G15.01-0.71, have been measured by \cite{Urquhart2018} at values of 2.30, 1.25, 2.97, and 1.86\,$\times$\,10$^5$\,$\rm L_\odot$, respectively. Assuming that the embedded young stellar objects (YSOs) or YSO candidates are the primary energy sources in these clumps based on the luminosity reported by \cite{Urquhart2018}, gas temperatures derived from the modified Stefan-Boltzmann blackbody radiation law were found to be 74.5, 65.9, 78.4, and 71.4\,K at radii of 0.12\,pc, respectively, consistent with the 51--71\,K range obtained from H$_{2}$CO line ratio measurements (see Table\,\ref{table:Clump-Parameters}). 
At larger radii of 0.24\,pc, the derived gas temperatures are 56.4, 50.0, 59.4, and 54.1\,K, respectively, aligning with the typical temperature of 50--60\,K derived from H$_{2}$CO in the M17SW region.
These results suggest that the protostars and/or YSOs embedded within the cloud emit significant radiation, effectively heating the surrounding molecular gas. Therefore, internal radiation may play a substantial role in gas heating within the M17SW region on a small scale.

As previously discussed, the complex temperature structure of the M17SW region may be attributed to both large-scale external radiative heating and small-scale internal radiative heating. The presence of a significant gas temperature gradient across the M17SW region, as measured by H$_2$CO and NH$_3$, provides evidence that the dense gas is primarily heated by radiation emitted from OB star clusters within the cavity. On a smaller scale, the dense gas surrounding the dense clumps experiences significant heating from internal protostars and/or YSOs. However, the spatial resolution of our observations, approximately 0.12\,pc, does not allow for a detailed measurement of the temperature structures within these dense clumps. Therefore, future high-resolution observations using ALMA will be necessary to overcome this limitation and provide a more comprehensive understanding of the temperature distribution in the region.

\subsection{Turbulent heating} 
\label{Sect:Turbulent-heating}
Previous observations of the M17SW region have revealed a complex velocity structure \citep{Hobson1992} with turbulence playing a significant role in broadening spectral lines (e.g., \citealt{Guesten1988,Perez-Beaupuits2015a,Hoang2022}). \cite{Guesten1988} found a strong correlation between local turbulence and gas temperature measured by NH$_{3}$\,(1,1) and (2,2) lines on a scale of approximately 0.4\,pc, indicating that dense gas in the M17SW region is heated by turbulence. 

Our study also addresses the relationship between turbulence and kinetic temperature on a scale of unprecedented $\sim$0.2\,pc.
To quantitatively analyze turbulence, we utilized the non-thermal line width ($\sigma_{\rm NT}$) of H$_{2}$CO. Positions with Mach numbers $\mathcal{M}\,\geq$\,4, located near dense regions significantly impacted by non-thermal motions, were selected. The correlation between non-thermal line width and kinetic temperature for both H$_{2}$CO\,3$_{03}$--2$_{02}$ and the combined H$_{2}$CO\,3$_{22}$--2$_{21}$ and 3$_{21}$--2$_{20}$ lines is depicted in Fig.\,\ref{fig:NT-T}. The correlation was fitted with a power law, showing $T_{\rm kin}\propto\sigma_{\rm NT}^{0.47\pm0.06}$ for H$_{2}$CO\,3$_{03}$--2$_{02}$ and $T_{\rm kin}\propto\sigma_{\rm NT}^{0.55\pm0.04}$ for the combined H$_{2}$CO\,3$_{22}$--2$_{21}$ and 3$_{21}$--2$_{20}$ lines. The correlation coefficients, $r$, were determined to be 0.30 and 0.52, respectively. These results suggest that the observed temperature increase in regions with strong non-thermal motions can be attributed to turbulent activities occurring on scales of $\sim$0.2\,pc, consistent with previous observational findings \citep{Guesten1988}.
The fitted results of H$_{2}$CO\,3$_{03}$--2$_{02}$ and the combined H$_{2}$CO\,3$_{22}$--2$_{21}$ and 3$_{21}$--2$_{20}$ lines in the M17SW region align with previous findings in the massive star-forming regions OMC-1 and DR21 \citep{Tang2018a,Zhao2024} and other massive clumps (ATLASGAL sample, \citealt{Tang2018b}) on scales ranging from 0.06 to 2\,pc. 
The high-density environment of M17SW ($n$(H$_2$)\,=\,5.5$\times$10$^{5}$\,cm$^{-3}$) promotes rapid radiative cooling 
through atomic fine-structure lines such as [OI], [CI], and [CII], 
the latter having been previously detected in this region \citep{Keene1985,Genzel1988,Stutzki1988,Perez-Beaupuits2010}. 
Consequently, the slightly lower power-law index of the H$_2$CO\,3$_{03}$--2$_{02}$ transition likely reflects a regime in which turbulent energy injection is largely balanced by efficient radiative cooling. 
While turbulence acts as a heating source, enhanced density contrasts and velocity gradients associated with turbulent motions may increase the effectiveness of line cooling, preventing a strong rise in gas temperature with increasing linewidth. 
As a result, the positive temperature–turbulence correlation observed in other star-forming regions is significantly weakened.

\subsection{Comparison with the Orion Bar}
\label{Sect:comparison}

Due to its proximity ($\sim$400\,pc, \citealt{Menten2007,Kounkel2017}), the Orion Bar is recognized as one of the clearest PDRs of our Galaxy. Gas temperature structures within the Orion Bar have been elucidated through observations conducted using the APEX\,12\,m and IRAM\,30\,m telescopes, where the same H$_{2}$CO\,(3--2) lines were mapped at scales of approximately 0.06 and 0.024\,pc, respectively \citep{Leurini2010,Tang2018a}. High temperatures ranging from 100 to 130\,K are detected in the dense gas at the edge of the Orion Bar, while lower temperatures of around 73\,K are observed within the internal dense clumps of the Orion Bar \citep{Tang2018a}. 
Notably, there is no evidence indicating massive star formation within the Orion Bar region. 
This may be attributed to its lower H$_{2}$ density and less effective shielding compared to M17SW, where denser gas and more efficient shielding create more favorable conditions for star formation.
Gas heating in the Orion Bar is proposed to be primarily influenced by far ultraviolet (FUV) photons emitted from the Trapezium stars located in the northern  H\,{\scriptsize II} region \citep{Tang2018a}. In contrast, the M17SW region exhibits prominent internal star formation activities, potentially leading to a more complex gas heating scenario in this region. On a larger scale, dense gas heating in the M17SW region may be significantly impacted by FUV photons originating from OB star clusters within the M17 cavity. Conversely, in smaller regions surrounding embedded protostars or clusters, dense gas heating likely results from a combination of processes including star formation activity, radiation, and turbulence, presenting a notable distinction from the conditions observed in the Orion Bar region.

\subsection{Influence of star formation}
\label{Sect:jeans-stability}
Star formation can be triggered by feedback from newly born massive OB stars, a mechanism thoroughly explained by the radiation-driven implosion model. The skin of the molecular cloud is heated and ionized by the photoionizing radiation of the 
nearby OB stars, and the high pressure drives a reverse shock by the rocket effect that propagate through to the 
cool interior of the clumps (e.g., \citealt{Dale2012,Nakatani2019}). The FUV photons dissociate 
and heat the molecular gas, and affect the physics and chemistry of gas regulating star formation.
FUV radiation may limit the final mass of the star through the photoevaporation of the outer layers of collapsing isothermal 
spheres which may rapidly deplete the envelope material, and trigger star formation with the compression of noncollapsing 
clumps by shock waves from the warm surface gas \citep{Gorti2002}.
High gas temperature will affect the Jeans masses of dense cores (Jeans' mass $M_{\rm Jeans}$\,=\,$1.25 \times (\frac{T_{\rm kin}}{10\,\rm K})^{3/2} \times (\frac{n(\rm H_2)}{10^5\,\rm cm^{-3}})^{-1/2} ~{\rm M_\odot}$, where $n(\rm H_2)$ is volume density and $T_{\rm kin}$ is the gas kinetic temperature) and may affect the initial mass function (IMF) of star formation.
The Jeans mass is about 7\,$\rm M_\odot$ for the mean physical parameters of the molecular gas in M17SW ($T_{\rm kin}$\,$\sim$\,54\,K and $n$(H$_2$)\,$\sim$\,5.5$\times$10$^{5}$\,cm$^{-3}$). The gas temperatures in M17SW exceed values found in typical Solar Neighborhood clouds ($T_{\rm kin}$\,$\sim$\,10--16\,K; e.g., \citealt{Pineda2026}) and infrared dark clouds ($T_{\rm kin}$\,$\sim$\,10--20\,K; e.g., \citealt{Pillai2006,Chira2013}) by a factor $\sim$3--6, so that $M_{\rm Jeans}$ is lower by a factor $\sim$4--13 in these clouds for fixed density $n(\rm H_2)$. This indicates that the star-forming clumps in the M17SW region are massive (e.g., \citealt{Urquhart2018,Eden2019}), potentially facilitating the formation of high-mass stars and leading to a top-heavy IMF in the M17SW area.

\section{Summary}
\label{Sect:Sumamry}

The kinetic temperature structure of the PDR M17SW has been investigated with 
the H$_{2}$CO triplet ($J_{\rm K_aK_c}$\,=\,3$_{03}$--2$_{02}$, 3$_{22}$--2$_{21}$, 
and 3$_{21}$--2$_{20}$) transitions near 218\,GHz on a linear scale of $\sim$\,0.18\,pc (18\arcsec) 
with the IRAM 30\,m telescope. The main results are the following:

\begin{enumerate}
\item
The H$_{2}$CO\,3$_{03}$--2$_{02}$ emission shows an extended distribution and reveals a clumpy structure of the dense gas in M17SW.

\item
We determined the kinetic temperature by modeling the measured H$_2$CO\,0.5$\times$[(3$_{22}$--2$_{21}$\,+\,3$_{21}$--2$_{20})$/3$_{03}$--2$_{02}$] line ratios 
with the RADEX non-LTE radiative transfer model. The derived kinetic temperatures ranges from 28 to 181\,K with an 
average of 54.2\,$\pm$\,0.3\,K derived assuming a typical density of 5.5$\times$10$^{5}$\,cm$^{-3}$. In comparison to temperature 
measurements obtained from multiple transitions of NH$_3$\,(1,1)--(6,6) and FIR wavelengths,
which have lower angular resolution than H$_{2}$CO, the H$_2$CO lines exhibit 
temperatures similar to those measured by NH$_3$ but slightly higher than the values derived from FIR observations.

\item
The kinetic temperature measurements using H$_2$CO indicate that the highest temperature ($T_\mathrm{kin}$>100\,K) is associated 
with the ultra-compact H\,{\scriptsize II} region UC1 and the eastern edge of M17SW. Elevated temperatures (>70\,K) are observed at various locations 
including the eastern edge of the primary structure of M17SW, the edge of clump G15.01-0.71, clumps G15.03-0.67 and G14.99-0.70,
southern regions of clumps G15.03-0.65, G15.02-0.68, G15.01-0.67, and G15.01-0.69, as well as in proximity to H$_2$O and CH$_3$OH
masers. Moderate temperatures ranging from 50 to 70\,K are commonly found in the periphery of the primary structure of M17SW. 
Lower temperatures (<50\,K) are identified at the western edge of M17SW and the edges of clumps G15.01-0.71 and G14.99-0.70. 

\item
The temperature gradient within the M17SW region as indicated by the H$_2$CO\,(3--2) and NH$_3$\,(2,2)/(1,1) line ratios 
indicates that the dense gas in M17SW is strongly affected by the radiation emitted from the OB star cluster NGC\,6618 
in the cavity with temperatures decreasing from northeast to southwest.

\item
On a smaller scale, the protostars and/or YSOs embedded within the cloud emit substantial radiation, which effectively heats
the surrounding molecular gas. This internal radiation likely plays a significant role in the heating of dense gas within the 
M17SW region.

\item
The non-thermal line widths of H$_2$CO exhibit a correlation with the gas kinetic temperatures in the M17SW region, which indicates that higher temperatures measured by H$_2$CO are linked to turbulence on a scale of $\sim$0.2\,pc.

\end{enumerate}

The kinetic temperature of the molecular gas is a fundamental parameter in our attempts to comprehend the ISM. 
M17SW serves as an optimal location to investigate the interstellar medium and star formation within dense PDRs. 
Future studies with high-resolution observations will concentrate on this region to further utilize formaldehyde as a thermometer, 
enhancing our understanding of gas heating and its effects on the star formation process.

\begin{acknowledgements}
The authors thank the anonymous referee for helpful comments.
We thank the staff of the IRAM telescope for their assistance in observations. 
This work acknowledges the support of the National Key R\&D Program of China 
under grant Nos.\,2023YFA1608002 and 2022YFA1603103, the Chinese Academy of Sciences 
(CAS) “Light of West China” Program under grant No.\,xbzg-zdsys-202212, 
and the Tianshan Talent Training Program of Xinjiang Uygur Autonomous Region under grant Nos.\,2022TSYCLJ0005. 
It was also partially supported by the Tianshan Talent Training Program of 
Xinjiang Uygur Autonomous Region under grant No.\,2024TSYCTD0013, 
the National Natural Science Foundation of China under grant Nos.\,12173075, 12373029, 12403033, and 12463006, 
the Xinjiang Key Laboratory of Radio Astrophysics under grant No.\,2023D04033, 
the Natural Science Foundation of Xinjiang Uygur Autonomous Region 
under grant Nos.\,2022D01A359 and 2023D01A11, the Central Guidance for Local Science
and Technology Development Fund under grant No.\,ZYYD2025ZY23, and the Youth Innovation 
Promotion Association CAS.  C.\,Henkel acknowledges support by the Chinese Academy of 
Sciences President's International Fellowship Initiative under grant No.\,2025PVA0048. 
T.\,Liu, K.\,Wang, X.\,P.\,Chen, and J.\,W.\,Wu acknowledge support by the 
Tianchi Talent Program of Xinjiang Uygur Autonomous Region. 
This research has used NASA's Astrophysical Data System (ADS). 
\end{acknowledgements}

\bibliographystyle{aa} 
\bibliography{bibfile} 

\begin{appendix} 
\onecolumn

\noindent
\begin{minipage}[t]{0.48\columnwidth}

\section{Previous measurements of temperature in M17SW region}
\label{Sect:Previous}

Based on previous observations of low-$J$ transitions of $^{12}$CO, $^{13}$CO, and CH$_{3}$CCH, the temperature 
in the M17SW cloud core has been estimated to be between 50 and 60\,K  
(e.g., \citealt{Bergin1994,Wilson1999,Howe2000,Snell2000}). In contrast, the mean temperature of the overall 
cloud is approximately 30 to 35\,K. Observations of mid- to high-$J$ $^{12}$CO transitions reveal the presence
of warmer gas in the M17SW region (e.g., \citealt{Harris1987,Stutzki1988,Perez-Beaupuits2010}). The kinetic 
temperature of the warm dense gas ($n$(H$_2$)\,$\gtrsim$\,10$^{4}$\,cm$^{-3}$) is found to be 
at least 100\,K. A non-LTE estimate of the ambient conditions indicates that the high-density clumps (with densities around 5$\times$10$^{5}$\,cm$^{-3}$) have temperatures of $\lesssim$230\,K \citep{Perez-Beaupuits2010}. 
In contrast, temperatures in the interface between the molecular cloud and the H\,{\scriptsize II} region 
range from 200 to 500\,K. \cite{Guesten1988} utilized the Effelsberg 100\,m telescope (beam size $\sim$\,40$''$) 
to map the temperature structure of the M17SW region with multiple transitions of NH$_{3}$\,(1,1)--(6,6). Their results 
indicate that temperatures within the M17SW molecular cloud range from $\sim$30 to 275\,K, with a bulk 
temperature around 50\,K. The kinetic temperature decreases gradually to about 30\,K as one moves away 
from the radiating surface. Subsequent mapping of the M17SW molecular cloud was conducted using the Green Bank Telescope (GBT), 
focusing on the NH$_{3}$\,(1,1) and (2,2) lines (beam size $\sim$\,32$''$; \citealt{Keown2019}). The kinetic 
temperature in M17SW was measured using the line ratios of the NH$_{3}$\,(1,1) and (2,2) transitions, 
revealing a range of 20-->80\,K, with an average of $\sim$40\,K. These measurements of gas temperature 
in M17SW are listed in Table\,\ref{table:M17SW-previous-T}.

The dust temperature structure of the M17SW region has been estimated in the wavelength range of 30 to 850\,$\mu$m 
(e.g., \citealt{Gatley1979,Meixner1992,Dupac2002,Hoang2022}). The analysis indicates that the dust 
temperature varies uniformly across the M17SW region. Using the dust emission from 30 and 100\,$\mu$m, 
\cite{Gatley1979} derived a color temperature map showing that temperatures in the M17SW cloud decrease 
from over 120\,K to below 50\,K as the distance from the H\,{\scriptsize II} region increases. 
\cite{Dupac2002} measured the dust temperature using emission data from 200 to 580\,$\mu$m, 
revealing that temperatures exceeding 80\,K (around 100\,K or higher) in regions near the ionization front, 
while $T_\mathrm{\rm dust}$ is 29\,K at the intensity peak of M17SW. Additionally, 
\cite{Hoang2022} measured the dust temperature utilizing {\it Herschel} data from 160 to 500\,$\mu$m and 
SCUBA2 data at 850\,$\mu$m. Their dust temperature map shows a decrease from over 63\,K at the northeast 
edge to below 30\,K at the southwest edge of M17SW, also emphasizing a gradient in dust temperature consistent 
with that reported by \cite{Dupac2002}. These measurements of dust temperature in M17SW are detailed in
Table\,\ref{table:M17SW-previous-T}.

\end{minipage}
\hfill
\begin{minipage}[t]{0.50\textwidth}


\captionof{table}{Previous measurements of temperature in M17SW.}
\centering
\rotatebox{90}{
\begin{tabular}
{llcccl}
\hline\hline
Molecule/ wavelength & Telescope & Beam size &$T$(range) & $T$(average) & References  \\
 &  &"  &K &K & \\
\hline
NH$_{3}$\,(1,1)--(6,6) & Effelsberg 100m & 40& 30--275 & $\sim$50 &\cite{Guesten1988} \\
$^{12}$CO\,1--0 & FCRAO 14m  & 45 & 26--62 & 48 & \cite{Bergin1994} \\
CH$_{3}$CCH\,6--5 & FCRAO 14m  & 50 & 25--52 & 34 & \cite{Bergin1994} \\
$^{13}$CO\,1--0\,\&\,5--4 & FCRAO 14m\,\&\,SWAS & 47 \& 240 &  20--63 &35 & \cite{Howe2000} \\
$^{12}$CO\,7--6\,\&\,6--5 & APEX & 7.7 &40--$\lesssim$230 & ... & \cite{Perez-Beaupuits2010} \\
NH$_{3}$\,(1,1)--(2,2) & GBT      & 32& 20-->80 & $\sim$40 & \cite{Keown2019}      \\
\hline
30, 50, 100\,$\mu$m & KAO & 60 & 50--100 & ... & \cite{Gatley1979} \\
50, 100\,$\mu$m & KAO & 30\,\&\,40 & $<$40--$>$75 & ... & \cite{Meixner1992} \\
200, 260, 360, 580\,$\mu$m & PRONAOS & 210 & 10--$>$80 & ...  & \cite{Dupac2002} \\
160, 250, 350, 500\,$\mu$m & {\it Herschel} & 36 & $<$30--$>$63 & $\sim$43 & \cite{Hoang2022} \\
\hline
\end{tabular}}
\label{table:M17SW-previous-T}
\end{minipage}

\clearpage

\noindent

\begin{minipage}{\textwidth}
\section{H$_{2}$CO velocity channel maps}
  \centering
  \includegraphics[width=0.7\textwidth]{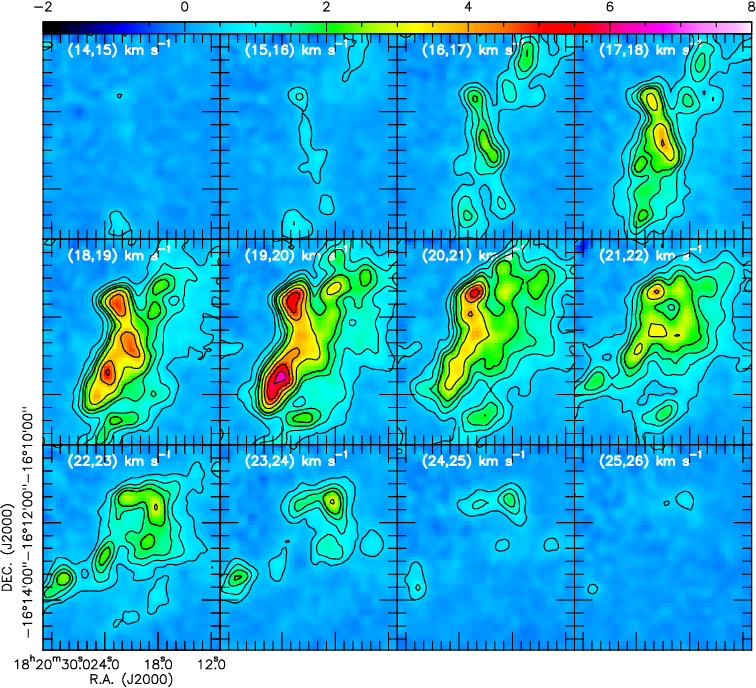}
  \captionof{figure}{Channel maps of the H$_{2}$CO\,(3$_{03}$--2$_{02}$) transition. The contour levels are running from 0.5 to 2.0\,K\,km\,s$^{-1}$ in steps of 
0.5\,K\,km\,s$^{-1}$ and from 2 to 8\,K\,km\,s$^{-1}$ in steps of 1\,K\,km\,s$^{-1}$ ($T_{\rm mb}$ scale; color bar in units of K\,km\,s$^{-1}$).}
\label{fig:H2CO-channel}
\end{minipage}

\vspace{-0.3cm}

\section{The LTE and non-LTE models for H$_{2}$CO and NH$_{3}$ temperature map}

\noindent

\makebox[\textwidth]{
\begin{minipage}[t]{0.48\textwidth}
  \centering
  \includegraphics[width=0.65\textwidth]{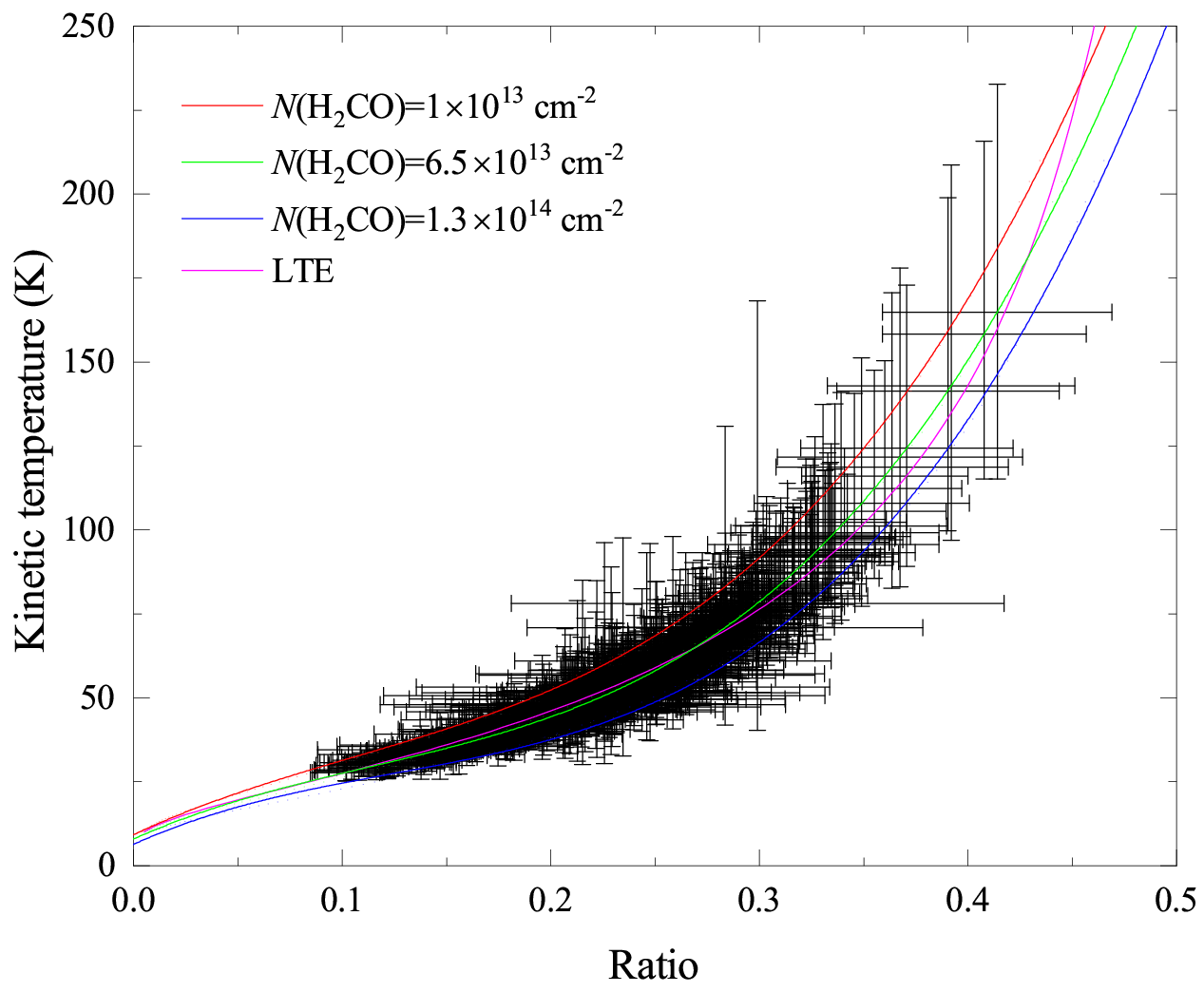}
  \captionof{figure}{The relation between the kinetic temperature and the average ratio of
H$_{2}$CO\,3$_{22}$--2$_{21}$/3$_{03}$--2$_{02}$ and 3$_{21}$--2$_{20}$/3$_{03}$--2$_{02}$ modeled by RADEX non-LTE with 
an assumed density of $n$(H$_2$)\,=5.5$\times$10$^{5}$\,cm$^{-3}$, column densities $N$(para-H$_{2}$CO)\,=1.0$\times$10$^{13}$,
6.5$\times$10$^{13}$ and 1.3$\times$10$^{14}$\,cm$^{-2}$ (\emph{red}, \emph{green}, and \emph{blue}), and an averaged linewidth of 3.8\,km\,s$^{-1}$.
The LTE kinetic temperature ($T_{\rm LTE}$, \emph{magenta}) is calculated as $T_{\rm LTE}\,=\,\frac{47.1}{{\rm ln}(0.556/{\rm ratio})}~{\rm K}$ (where the ratio represents the average ratio of H$_{2}$CO\,3$_{22}$--2$_{21}$/3$_{03}$--2$_{02}$ and
3$_{21}$--2$_{20}$/3$_{03}$--2$_{02}$) using the methodology described in \cite{Mangum1993} and \cite{Tang2017b}. 
The black points were derived from observed H$_{2}$CO line ratios  
in the M17SW region for a column density $N$(para-H$_{2}$CO)\,=\,6.5$\times$10$^{13}$\,cm$^{-2}$. Temperature uncertainties were determined based on observed H$_{2}$CO 
line ratio errors.}
  \label{fig:Radex}
\end{minipage}
\hfill

\begin{minipage}[t]{0.48\textwidth}
  \centering
  \includegraphics[width=0.5\textwidth]{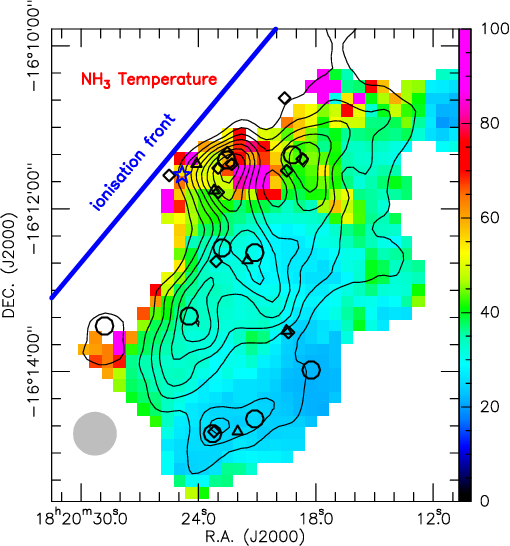}
  \captionof{figure}{Map of the kinetic temperatures obtained from NH$_{3}$\,(2,2)/(1,1) line ratios observed with the GBT (also see Fig.11 in \citealt{Keown2019}), represented by the color bar in units of Kelvin.
Black contours show the integrated intensity of H$_{2}$CO\,3$_{03}$--2$_{02}$ (same as in Fig.\,\ref{fig:H2CO-intensity-maps}).
The marks are the same as in Fig.\,\ref{fig:H2CO-intensity-maps}. The beam size of the NH$_{3}$ data is shown in the lower left corner.
The blue solid line indicates the location of the assumed ionization front.}
  \label{fig:NH3_T}
\end{minipage}
}

\end{appendix}

\end{document}